\documentclass[paper,11pt]{JHEP}
\usepackage[centertags]{amsmath}
\usepackage{amsfonts} \usepackage{amssymb} \usepackage{amsthm}
\allowdisplaybreaks[1]
\usepackage{graphicx}
\newcommand{\Z}{{\mathbb Z}}

\newcommand{\C}{{\mathbb C}}
\newcommand{\eq}[1]{Eq.~(\ref{#1})}

\newcommand{\ii}{\mathrm{i}}
\newcommand{\ee}{\mathrm{e}}

\newcommand{\vev}[1]{\langle #1 \rangle}

\newcommand{\diag}{\mathrm{diag}\,}
\newcommand{\cA}{{\mathcal{A}}}

\newcommand{\cD}{{\mathcal{D}}}
\newcommand{\cE}{{\mathcal{E}}}

\newcommand{\cL}{{\mathcal{L}}}

\newcommand{\cN}{{\mathcal{N}}}
\newcommand{\cO}{{\mathcal{O}}}

\newcommand{\one}{{\rm 1\kern -.9mm l}}
\newcommand{\bone}{\mathbf{1}}
\newcommand{\ft}[2]{{\textstyle\frac{#1}{#2}}}

\newcommand{\eud}{\epsilon_1\epsilon_2}

\newdimen\tableauside\tableauside=1.0ex
\newdimen\tableaurule\tableaurule=0.4pt
\newdimen\tableaustep
\def\phantomhrule#1{\hbox{\vbox to0pt{\hrule height\tableaurule
width#1\vss}}}
\def\phantomvrule#1{\vbox{\hbox to0pt{\vrule width\tableaurule
height#1\hss}}}
\def\sqr{\vbox{%
  \phantomhrule\tableaustep
\hbox{\phantomvrule\tableaustep\kern\tableaustep\phantomvrule\tableaustep}%
  \hbox{\vbox{\phantomhrule\tableauside}\kern-\tableaurule}}}
\def\squares#1{\hbox{\count0=#1\noindent\loop\sqr
  \advance\count0 by-1 \ifnum\count0>0\repeat}}
\def\tableau#1{\vcenter{\offinterlineskip
  \tableaustep=\tableauside\advance\tableaustep by-\tableaurule
  \kern\normallineskip\hbox
    {\kern\normallineskip\vbox
      {\gettableau#1 0 }%
     \kern\normallineskip\kern\tableaurule}%
  \kern\normallineskip\kern\tableaurule}}
\def\gettableau#1 {\ifnum#1=0\let\next=\null\else
  \squares{#1}\let\next=\gettableau\fi\next}
\newcommand{\Yfund}{\tableau{1}}
\newcommand{\Ysymm}{\tableau{2}}
\newcommand{\Yasymm}{\tableau{1 1}}

\def\XXint#1#2#3{{\setbox0=\hbox{$#1{#2#3}{\int}$}
     \vcenter{\hbox{$#2#3$}}\kern-.5\wd0}}

\title{S-duality and the prepotential in $\mathcal{N}=2^\star$ theories (I): the ADE algebras
}
\author{M. Bill\'o$^1$, M. Frau$^{1}$, F. Fucito$^{2}$, A. Lerda$^{3,1}$, J.F. Morales$^{2}$
\\
\vskip 0.2cm
$^1$ Universit\`a di Torino, Dipartimento di Fisica
\\ and I.N.F.N. - sezione di Torino,
Via P. Giuria 1, I-10125 Torino, Italy\\
\vskip 0.2cm
$^2$ I.N.F.N - sezione di Roma 2\\
and Universit\`a di Roma Tor Vergata, Dipartimento di Fisica\\
Via della Ricerca Scientifica, I-00133 Roma, Italy
\vskip 0.2cm
$^3$Universit\`a del Piemonte Orientale, Dipartimento di Scienze e Innovazione Tecnologica, \\
and I.N.F.N. - Gruppo Collegato di Alessandria - sezione di Torino\\
Viale T. Michel  11, I-15121 Alessandria, Italy\\
\vspace{0.35cm}
\email{billo,frau,lerda@to.infn.it; fucito,morales@roma2.infn.it}
}
\abstract{The prepotential of $\mathcal{N}=2^\star$ supersymmetric theories with unitary gauge groups
in an $\Omega$ background satisfies a modular anomaly equation that can be recursively solved order 
by order in an expansion for small mass. By requiring that S-duality acts
on the prepotential as a Fourier transform we generalise this result to  $\mathcal{N}=2^\star$ theories with gauge algebras of the D and E type and show that
their prepotentials can be written in terms of quasi-modular forms of $\mathrm{SL}(2,\mathbb{Z})$. The results are checked against microscopic multi-instanton calculus based on localization for the A and D series
and reproduce the known 1-instanton prepotential of the pure $\mathcal{N}=2$ theories for 
any gauge group of ADE type.
Our results can also be used to obtain the multi-instanton terms in  the exceptional theories for 
which the microscopic instanton calculus and the ADHM construction are not available.
}


\preprint{ROM2F/2015/7}

\begin{document}
\section{Introduction}
\label{secn:intro}

Gauge theories with extended supersymmetries show very remarkable behaviours.
For example the maximally supersymmetric Yang-Mills theories in $d=4$, briefly $\cN=4$ SYM theories,
are conjectured to enjoy S-duality invariance.  S-duality is a strong/weak-coupling relation that
exchanges electrically charged states  with non-perturbative magnetically charged
states \cite{Montonen:1977sn}; over the years many tests of this conjecture have been 
carried out with success (see for instance \cite{Vafa:1994tf,Girardello:1995gf}).
For theories with simply-laced gauge groups, S-duality maps the gauge group to itself, but
when the gauge group is not simply laced, the gauge group of the S-dual theory is
the GNO dual group \cite{Goddard:1976qe}.

Also $\cN=2$ SYM theories are very interesting: even if they are less constrained than the $\cN=4$ theories,
it is still possible to study several of their properties in an exact way.
Indeed their perturbative contribution is exhausted at the one-loop level and
their non-perturbative behaviour is by now well understood, on the one hand, via the Seiberg-Witten \cite{Seiberg:1994rs,Seiberg:1994aj} description of their low energy effective theory and, on the other hand, via the direct computation of instanton effects by means of localization techniques \cite{Nekrasov:2002qd}\nocite{Flume:2002az,Nekrasov:2003rj,Bruzzo:2002xf}\,-\,\cite{Fucito:2004ry}.
A noticeable exception in this scenario is given by theories with exceptional gauge groups for which an ADHM construction of the instanton moduli space is still missing%
\footnote{See for example \cite{Gaiotto:2012uq}\nocite{Keller:2012da,Benvenuti:2010pq,Keller:2011ek,Hanany:2012dm}\,-\,\cite{Cremonesi:2014xha} for recent progresses on the description of instanton moduli spaces in theories with exceptional gauge groups}.

Among the $\cN=2$ models much attention has been devoted, in  the recent years, to superconformal
theories and to their mass deformations, which sit at the crossroad
of many approaches to the non-perturbative description of quantum field theories and of their duality structures (see for example the collective review \cite{Teschner:2014oja} and the references therein).
This paper deals with the $\cN=2^\star$ SYM theories with simply laced gauge group $G$ whose 
corresponding Lie algebra will be denoted by $\mathfrak{g}$.
Beside the $\cN=2$ gauge vector multiplet, these theories contain
an adjoint hypermultiplet of mass $m$ and represent a mass deformation of the $\cN=4$ SYM theory.
In an appropriate large-$m$ limit, the hypermultiplet decouples and the pure $\cN=2$ SYM theory is retrieved.
The $\cN=2^\star$ theory inherits from the $\cN=4$ theory an interesting action of S-duality. In particular,
S-duality acts non-trivially on the prepotential function $F$ that encodes
the low-energy effective dynamics on the Coulomb branch of moduli space.
Upon expanding the prepotential in powers of the mass $m$, this action can be exploited
to efficiently determine the non-perturbative expression of the prepotential.
This is achieved by showing that the coefficients $f_n$ of the
mass expansion of $F$ are (quasi)-modular functions of the gauge coupling $\tau$ connected
to each other by a recursion relation.

Such a recursion relation, which encodes the ``modular anomaly'' of the prepotential,
was first pointed out for U$(N)$ theories in \cite{Minahan:1997if} where it was derived from
the Seiberg-Witten curve.
The modular anomaly is related to the holomorphic anomaly of topological string amplitudes through
local Calabi-Yau embeddings of the SW curves \cite{Bershadsky:1993ta}\nocite{Witten:1993ed,Aganagic:2006wq}\,-\,\cite{Gunaydin:2006bz}.
It has been studied also in presence of an $\Omega$-background
\cite{Huang:2006si}-\nocite{Grimm:2007tm,Huang:2009md,Huang:2010kf,Huang:2011qx,Galakhov:2012gw,Billo:2013fi,Billo:2013jba,Nemkov:2013qma,Billo:2014bja}\cite{Lambert:2014fma}, in the framework of the AGT correspondence
\cite{Marshakov:2009kj}-\nocite{KashaniPoor:2012wb,Kashani-Poor:2013oza}\cite{Kashani-Poor:2014mua}
and in $\cN=2$ conformal SQCD models with fundamental matter 
\cite{Billo:2013fi,Billo:2013jba,Ashok:2015cba}.

Here we review and streamline the derivation of the modular anomaly equation and
the associated recursion relation directly from the S-duality requirement and for a generic
simply-laced gauge group $G$ (the non simply-laced groups will be discussed
in a companion paper \cite{Billo}).
The modular anomaly equation leads to express the prepotential in terms of modular
forms of $\tau$ and of functions of the periods $a$
which are written in terms of the root system of $\mathfrak{g}$, allowing for a unified treatment of all Lie algebras.
In this way we can compute the prepotential also for the exceptional Lie algebras $E_6$, $E_7$ and $E_8$
for which an ADHM construction of the instanton moduli space is not available.

This is the plan of the paper: in Section~\ref{secn:sduality} we discuss the behaviour of the
$\mathcal{N}=2^\star$ theories under S-duality and
derive from it the modular anomaly equation satisfied by the prepotential.
In Section~\ref{secn:recursion} we exploit the recursion relation equivalent to the modular anomaly equation to
compute exactly, {\it{i.e.}} to all orders in the instanton expansion, the first few coefficients
in the mass expansion of the prepotential. The results are then
generalised in Section~\ref{sec:recepsi} to account for a non-trivial $\Omega$-background.
In Section~\ref{snek} we will describe the direct microscopic computation of the instanton corrections for the algebras of type $A_r$ and $D_r$ using the equivariant localization methods. The purpose of this section
is to clarify some subtle points of the multi-instanton calculus and to check successfully these microscopic results against the instanton expansion of the solutions of the modular anomaly equation derived in the previous section. Our conclusions are presented in Section~\ref{sec:concl}, while several
technical material is confined in various appendices.

\section{S-duality and modular anomaly}
\label{secn:sduality}
In this section we briefly review the structure of $\cN=2^\star$ theories with a gauge group $G$ of ADE type and discuss the constraint that S-duality imposes on their prepotential.

\subsection{The $\mathrm{SL}(2,\mathbb{Z})$-duality symmetry}
The field content of these theories includes an  $\cN=2$ vector multiplet and a massive hypermultiplet, both transforming in the adjoint representation of $G$. The $\cN=2$ gauge multiplet contains an adjoint complex scalar $\varphi$, whose vacuum expectation value can always be aligned along the Cartan directions
and written in the diagonal form
\begin{equation}
\label{phivev}
\vev{\varphi} =a= \diag (a_1,a_2,\ldots,a_r)~,
\end{equation}
with $r$ denoting the rank of the gauge Lie algebra $\mathfrak{g}$. The parameters $\{ a_u \}$ span the Coulomb branch of the classical moduli space of the gauge theory.  The low energy effective action on this branch is specified by a holomorphic function: the prepotential $F(a)$.
Alternatively the gauge theory can be described in terms of the dual variables
\begin{equation}
\label{secada0}
a^{\mathrm{D}}_u = \frac{1}{2\pi\ii}\frac{\partial F}{\partial a_u}~.
\end{equation}
In the following we will often write $\partial_u$ for $\ft{\partial}{\partial a_u}$. We will also use a simplified vector notation, writing, for instance, $a$ for the vector, $\ft{\partial}{\partial a}$ for the gradient vector,
and so on.

The effective coupling matrix, which also encodes the metric on the moduli space, is
\begin{equation}
\label{tauuvdef}
\tau_{uv} = \partial_u a^{\mathrm{D}}_{v} = \frac{1}{2\pi\ii}\partial_u\partial_v F~.
\end{equation}
The classical part of the prepotential reads simply
\begin{equation}
\label{iprepclass}
F^{\mathrm{cl}}=  \pi\ii\tau a^2
\end{equation}
where $\tau$ is the complexified gauge coupling
\begin{equation}
\label{tau}
\tau=\frac{\theta}{2\pi}+\ii\frac{4\pi}{g^2}
\end{equation}
At this level, the dual periods and the effective coupling matrix are
\begin{equation}
\label{adclass}
a^{\mathrm{D}} = \tau a~,~~~   \tau_{uv} = \tau\,\delta_{uv}~.
\end{equation}
In a Seiberg-Witten description of the theory, $a$ and $a^{\mathrm{D}}$ describe the periods
of the Seiberg-Witten differential along the $2r$ cycles of the Riemann-surface
defined by the Seiberg-Witten curve.
The periods and dual periods can be assembled in a $2r$-dimensional vector
$\big(a^{\mathrm{D}},a\big)$ that transforms as a vector of the modular
group $\mathrm{Sp}(4r,\mathbb{Z})$ of the Riemann surface. The two set of variables are suitable to
describe the regimes of weak and strong coupling ($g$ small and $g$ large respectively) of the gauge theory.
These two regimes are mapped to each other by S-duality which, as an
element of $\mathrm{Sp}(4r,\mathbb{Z})$, exchanges periods and dual periods
and acts projectively on $\tau$ by inverting it, namely
\begin{equation}
\label{Saad}
S(a) = a^{\mathrm{D}}~,~~~S(a^{\mathrm{D}}) = - a~,~~~S(\tau) = -\frac{1}{\tau}~.
\end{equation}
On the other hand, the T-duality acts as
\begin{equation}
\label{Taad}
T(a) = a~,~~~T(a^{\mathrm{D}}) =  a^{\mathrm{D}}+a~,~~~T(\tau) = \tau+1 ~.
\end{equation}
S and T generate the SL$(2,\Z)$ modular group.

On the prepotential, the T-duality action is 
\begin{equation}
\label{Tduality}
T[F(a)]=F(a)+\pi \ii a^2~,
\end{equation}
as one can see from the fact that only the classical part $F^{\mathrm{cl}}$
given in (\ref{iprepclass}) transforms non-trivially under $\tau\to \tau+1$. Indeed,
$\cN=2$ supersymmetry allows only for one-loop ($\tau$-independent) and instanton corrections 
(weighted by $\ee^{2\pi \ii k \tau}$ with $k$ integer) which are $T$-invariant.

The S-duality action is instead much less trivial since it maps the description of the theory
in the variables $a$ to its dual
description in terms of $a^{\mathrm{D}}$. Therefore $S$ should map the prepotential $F(a)$ to
its Legendre transform:
\begin{equation}
\label{LTFcl}
S[ F(a) ] =\cL[F](a^{\mathrm{D}})
\end{equation}
where
\begin{equation}
\label{LTF}
\cL[F](a^{\mathrm{D}}) =  F(a) - 2\pi\ii \,a\cdot a^{\mathrm{D}} = F(a) - a\cdot\frac{\partial F}{\partial a}~.
\end{equation}
The classical part of the prepotential verifies immediately (\ref{LTFcl}); in fact
\begin{equation}
\label{SF}
S[F^{\mathrm{cl}}] = -\frac{\pi\ii}{\tau}\,\big(a^{\mathrm{D}}\big)^2
=\cL[F^{\mathrm{cl}}](a^{\mathrm{D}})~.
\end{equation}
The $S$-duality symmetry requirement (\ref{LTFcl}) represents instead a highly non-trivial constraint
on the \emph{quantum} prepotential. As we will see, this constraint allows us to determine
the exact form of the prepotential, order by order in the hypermultiplet mass, starting from
very few microscopic data.

As mentioned in the introduction, this is known to happen for U$(N)$ theories, where the prepotential satisfies a ``modular anomaly'' equation  that has been discussed extensively in the literature
\cite{Minahan:1997if}-\cite{Billo:2014bja}. In the following we derive the modular anomaly equation directly from the S-duality relation (\ref{LTFcl}) and show that it holds for gauge theories with all gauge groups of ADE type.

\subsection{The small mass expansion of the prepotential}
\label{subsec:n2}

When the mass $m$ of the adjoint hypermultiplet vanishes, there are no quantum corrections to the prepotential since the supersymmetry is enhanced to $\cN=4$.
When the mass is turned on, the supersymmetry is only $\cN=2$ and the prepotential $F$ is
corrected. We write
\begin{equation}
\label{iprep1}
F = F^{\mathrm{cl}} + f = \pi\ii\tau a^2 + f~,
\end{equation}
where $f$ is the quantum part of the prepotential.
The dual periods and effective coupling $\tau_{uv}$ also get
quantum corrected and become non-trivial functions of $\tau$.
As already mentioned, $\cN=2$ supersymmetry allows perturbative corrections only at one-loop and
non-perturbative corrections at all instanton numbers.
Working in a mass expansion, we write the quantum prepotential as
\begin{equation}
f = f^{\mathrm{1-loop}}+f^{\mathrm{inst}}=\sum_{n=1}^\infty f_n
\label{iprepfexp}
\end{equation}
where
\begin{equation}
\label{fn1linst}
f_n = f^{\mathrm{1-loop}}_n+f^{\mathrm{inst}}_n
\end{equation}
is proportional to $m^{2n}$.

The one-loop contribution to the prepotential has the form (see for instance \cite{D'Hoker:1999ft})
\begin{equation}
\label{F1loop}
\begin{aligned}
 f^{\mathrm{1-loop}}& = \frac{1}{4}\sum_{\alpha\in\Psi} \left[
 -(\alpha\cdot a)^2 \log\left(\frac{\alpha\cdot a}{\Lambda}\right)^2 +
 (\alpha\cdot a+ m)^2 \log\left(\frac{\alpha\cdot a+m}{\Lambda}\right)^2  \right]
 \end{aligned}
\end{equation}
where $\Lambda$ is an arbitrary scale and
$\alpha$ is an element of the root system $\Psi$ of the algebra $\mathfrak{g}$;
$\alpha$ is an $r$-dimensional vector of components $\alpha_u$. The scalar product
$\alpha\cdot a$
represents the mass acquired by the complex $W$-boson associated to the root $\alpha$ via the
(super)-Higgs mechanism. Also the mass of the adjoint scalar along the root $\alpha$ is shifted with respect to its original value $m$ by the same amount.
Expanding (\ref{F1loop}) in powers of $m$, all odd powers cancel upon summing over
positive and negative roots, and we find
\begin{equation}
\label{fn1loop}
\begin{aligned}
f^{\mathrm{1-loop}} & =  \frac{m^2}{4} \sum_{\alpha\in\Psi}  \log\left(\frac{\alpha\cdot a}{\Lambda}\right)^2
- \sum_{n=2}^\infty  \frac{m^{2n}}{4n(n-1)(2n-1)}\, C_{2n-2}\\
& = \frac{m^2}{4} \sum_{\alpha\in\Psi}  \log\left(\frac{\alpha\cdot a}{\Lambda}\right)^2 -\frac{m^4}{24} \,C_2
-\frac{m^6}{120}\, C_4  -\frac{m^8}{336}\, C_6 - \ldots
\end{aligned}
\end{equation}
where we defined
\begin{equation}
\label{defCn}
C_{n} = \sum_{\alpha\in\Psi} \frac{1}{(\alpha\cdot a)^n }~.
\end{equation}

The non-perturbative part of the prepotential receives contributions from the various instanton sectors, so $f^{\mathrm{inst}}$ is a series in the instanton weight
\begin{equation}
\label{defq}
q = \ee^{-\frac{8\pi^2}{g^2}+\ii\,\theta} = \ee^{2\pi\ii\tau}~.
\end{equation}
The term of order $q^k$ can be evaluated integrating over the moduli spaces of $k$-instantons by means of localization techniques when the gauge group $G$ is one of the classical matrix groups. This  excludes the exceptional groups $E_{6,7,8}$.
We will review this computation in Section~\ref{snek}. This procedure can in principle be carried out up to arbitrary order $k$; in practice, however, it is computationally rather intense.
It is important to observe that
\begin{equation}
\label{Finst1}
f^{\rm inst}_1 =0
\end{equation}
since instanton contributions start at order $m^4$. This can be seen by noticing that every mass
insertion soaks two of the eight instanton fermionic zero modes of the $\cN=4$ theory, so we need at least four powers of $m$ to get a non-trivial result.

\subsection{The modular anomaly equation}
\label{subsec:Sprep}

We now investigate the consequences of the S-duality relation (\ref{LTFcl}) on the
quantum prepotential $f$. First we observe that the prepotential has mass dimension 2, so on dimensional grounds all $f_n$ with $n\geq 2$ must be homogeneous functions of
degree $2-2n$ in $a$:
\begin{equation}
f_n(\lambda a)= \lambda^{2-2n}\,f_n(a)~.
\label{homogeneity}
\end{equation}
Moreover, they are non-trivial functions of $\tau$ expressed as Fourier series in $q$. We therefore use
the notation $f_n(\tau,a)$ to express this fact.

Let us now compute first the two sides of the duality relation (\ref{LTFcl}).
The Legendre transform of $F$ is
\begin{equation}
\label{legF}
\cL[F]   = F - a \cdot \frac{\partial F}{\partial a} = - \pi\ii\tau a^2 + f - a \cdot\frac{\partial f}{\partial a}~.
\end{equation}
On the other hand, using (\ref{Saad}), the S-transform of $F$ is
\begin{equation}
\label{Sonfexplicit}
S[F] = - \frac{\pi\ii}{\tau} \,\big(a^{\mathrm{D}}\big)^2+ f\big(-\ft{1}{\tau},a^{\mathrm{D}}\big)~,
\end{equation}
where, according to (\ref{secada0}),
\begin{equation}
\label{secada}
a^{\mathrm{D}} = \frac{1}{2\pi\ii}\,\frac{\partial F}{\partial a}=
\tau \left( a +\frac{1}{2\pi\ii\tau}\frac{\partial f}{\partial a}\right)~.
\end{equation}
Plugging (\ref{secada}) into (\ref{legF}), the S-duality relation  $S[F] =\cL[F]$ can be written in the form
\begin{equation}
\label{Sfeq}
f \big( -\ft{1}{\tau}, a^{\mathrm{D}} \big)   =  f ( \tau, a)  +\frac{1}{4\pi\ii\tau}
 \left( \frac{\partial f (\tau,a)}{\partial a} \right)^2~.
\end{equation}
{From} (\ref{fn1loop}) and (\ref{Finst1}) we notice that $f_1$ is independent of $\tau$ but dependent on $\Lambda$, so equation (\ref{Sfeq}) at order $m^2$ implies
\begin{equation}
\label{f1is}
f_1(  \tau a, S(\Lambda) ) = f_1 ( a,\Lambda)
\end{equation}
where we have allowed an action of $S$-duality on the scale $\Lambda$.
Using the explicit form of $f_1$, that is
\begin{equation}
\label{f1is0}
f_1(a,\Lambda) = f_1^{\rm 1-loop}(a,\Lambda)
= \frac{m^2}{4} \sum_{\alpha\in\Psi}  \log\left(\frac{\alpha\cdot a}{\Lambda}\right)^2~,
\end{equation}
we conclude that  $S(\Lambda)=\tau \Lambda$.

At higher orders in the mass expansion, the differential equation (\ref{Sfeq}) can be solved by taking
$f_n$, for $n\geq 2$, to be an $\mathrm{SL}(2,\Z)$ quasi-modular form of weight $2n-2$.
A basis of quasi-modular forms is given  by the set of Eisenstein series
$\{ E_2, E_4, E_6 \}$.  More precisely $E_4$ and $E_6$ are true modular forms of weight $4$ and $6$ respectively, so under S-duality they transform as
\begin{equation}
\label{SE46}
E_4\big(-\ft{1}{\tau}\big) = \tau^{4}\,E_4 (\tau) ~, ~~~   E_6\big(-\ft{1}{\tau}\big)
= \tau^{6}\,E_6 (\tau) ~.
\end{equation}
The $E_2$ series is instead a quasi-modular form of weight 2 because under S it gets shifted:
\begin{equation}
\label{SE2}
E_2\big(-\ft{1}{\tau}\big) = \tau^{2}\, \Big(E_2 (\tau) + \frac{6}{\pi\ii\tau}\Big)\,\equiv\,
\tau^{2}\, \big(E_2 (\tau) + \delta\big)~.
\end{equation}
Here we introduced the notation $\delta = \ft{6}{\pi\ii\tau}$ to avoid clutter in the subsequent formul\ae\,. 

We notice that all $\delta$-dependence should cancel in the duality relation since $f$ is only a function of $q$,
and that the quasi-modularity of $f_n$ is due entirely to its dependence on $E_2$. Indicating explicitly this dependence,  we have  (for $n\geq 2$ )
\begin{equation}
\label{Sfn}
 f_n\big(-\ft{1}{\tau}, a^{\mathrm{D}},E_2(-\ft{1}{\tau})\big) =\tau^{2n-2}\, f_n\big(\tau,  a^{\mathrm{D}},E_2+\delta\big)=   f_n\big(\tau,\ft{a^{\mathrm{D}}}{\tau} ,E_2+\delta\big)~.
\end{equation}
where in the last step we used  the homogeneity property (\ref{homogeneity}) of $f_n$. 
On the other hand we have
\begin{equation}
\label{Sfn1}
 f_1(a^{\mathrm{D}}, \tau \Lambda)  =  f_1\big(\ft{a^{\mathrm{D}}}{\tau}, \Lambda \big)~.
\end{equation}
Plugging (\ref{Sfn}) and (\ref{Sfn1}) into the left hand side of (\ref{Sfeq}), we find
\begin{eqnarray}
f\big(-\ft{1}{\tau}, a^{\mathrm{D}},E_2(-\ft{1}{\tau}),\tau\Lambda\big)
&=&f\big(\tau, \ft{a^{\mathrm{D}}}{\tau},E_2+\delta,\Lambda\big)\phantom{\Big|}\nonumber\\
&=&f\Big(\tau, a + \frac{\delta}{12}\frac{\partial f}{\partial a},E_2 + \delta, \Lambda\Big)\nonumber\\
&=&f (\tau,a,E_2, \Lambda)+ \delta\,\left[ \frac{\partial f}{\partial {E_2}} +\frac{1}{12}
\left(\frac{\partial f}{\partial a}\right)^2  \right](\tau,a,E_2,\Lambda)\nonumber\\
&& + \frac{\delta^2}{2}\,\left[\frac{\partial^2 f}{\partial E_2^2} + \frac{1}{144} \left(\frac{\partial f}{\partial a}\right)^2
\frac{\partial^2 f}{\partial a^2} + \frac{1}{6} \frac{\partial f}{\partial a}\cdot \frac{\partial^2 f}{\partial a \partial E_2}\right](\tau,a,E_2,\Lambda)\nonumber\\
&&+ \cdots{\phantom{\Big|}} \label{Sfn2}
\end{eqnarray}
where the dots stand for higher order terms in $\delta$. Comparing (\ref{Sfn2}) with the right hand side of (\ref{Sfeq}), one finds that at order $\delta$ the following modular anomaly equation has
to be satisfied
\begin{equation}
\label{recdiff}
\frac{\partial f}{\partial E_2} +\frac{1}{24} \left(\frac{\partial f}{\partial a}\right)^2=0~.
\end{equation}
It is straightforward to check that the conditions obtained at higher orders in $\delta$ follow
from this equation. For instance, the term in $\delta^2$ in (\ref{Sfn2}) is easily shown
to correspond to a further $E_2$-derivative of the modular anomaly equation (\ref{recdiff}); 
thus it vanishes, as requested by the comparison with the right hand side of (\ref{Sfeq}).

Summarizing, the S-duality symmetry requirement (\ref{LTFcl}) is satisfied if
the coefficients $f_n$ in the mass expansion of the quantum prepotential $f$ are quasi-modular
form of weight $2n-2$ and if $f$ satisfies the modular anomaly equation (\ref{recdiff}).

\section{The recursion relation}
\label{secn:recursion}

\subsection{The prepotential}

Expanding the quantum prepotential $f$ in mass powers according to (\ref{iprepfexp}), the requirement (\ref{recdiff}) turns into the relation
\begin{equation}
\label{rec}
\frac{\partial f_n}{\partial E_2} = -\frac{1}{24} \sum_{m=1}^{n-1}
\frac{\partial f_m}{\partial a}\cdot \frac{\partial f_{n-m}}{\partial a}
\end{equation}
which allows to recursively determine the $f_n$'s in terms of the  lower coefficients up to
$E_2$-independent functions. The $E_2$-independent part
can be fixed by using one-loop or lower-$k$ instanton data.
Actually, to the order we will consider here, the one-loop data will be enough.

Let us start by determining $f_2$ which, being a quasi-modular form of weight 2,
can only be proportional to $E_2$.
For $n=2$ the recursion relation (\ref{rec}) simply reads
\begin{equation}
\label{recn2}
\frac{\partial f_2}{\partial E_2} =
-\frac{1}{24} \frac{\partial f_1}{\partial a}\cdot \frac{\partial f_1}{\partial a}=
-\frac{m^4}{96} \sum_{\alpha,\beta\in\Psi} \frac{\alpha\cdot\beta}{(\alpha\cdot a)(\beta\cdot a)}
~,
\end{equation}
where in the second step we have used the expression (\ref{f1is}) for $f_1$.
The sum over the roots $\alpha,\beta\in\Psi$ can be rewritten as
\begin{equation}
\label{df1df1}
\sum_{\alpha,\beta\in\Psi} \frac{\alpha\cdot\beta}{(\alpha\cdot a)(\beta\cdot a)}
=
4\,\sum_{\alpha\in\Psi} \frac{1}{(\alpha\cdot a)^2} +
\sum_{\alpha\not= \pm \beta\in\Psi} \frac{\alpha\cdot\beta}{(\alpha\cdot a)(\beta\cdot a)}~.
\end{equation}
The first term corresponds to the cases $\alpha = \pm \beta$
and comes with an overall factor of 4 since for any ADE Lie algebra
all roots have length square 2: $\alpha\cdot\alpha=2$ (see Appendix~\ref{secn:roots} for details on the
root system of the ADE algebras).
In the second term of (\ref{df1df1}), for any $\beta\not= \alpha$ we have
either $\alpha\cdot\beta=\pm 1$ or $\alpha\cdot\beta=0$, and
so both $\beta$ and $-\beta$ give the same contribution. Therefore,
we can limit ourselves to sum over the roots $\beta\in\Psi(\alpha)$ where
\begin{equation}
\label{defDeltaalpha}
\Psi(\alpha) = \left\{\beta\in\Psi:~ \alpha\cdot\beta=1 \right\}~,
\end{equation}
and get
\begin{equation}
\label{defc11}
\sum_{\alpha,\beta\in\Psi} \frac{\alpha\cdot\beta}{(\alpha\cdot a)(\beta\cdot a)}
 =
4 \sum_{\alpha\in\Psi} \frac{1}{(\alpha\cdot a)^2} +
2 \sum_{\alpha\in\Psi}\sum_{\beta\in\Psi(\alpha)} \frac{1}{(\alpha\cdot\phi)(\beta\cdot\phi)}~.
\end{equation}
The first term is proportional to $C_2$ as one can see from (\ref{defCn}), while the second term
suggests to introduce a more general sum over the root lattice, namely
\begin{equation}
\label{defcnms}
C_{n;m_1m_2\ldots m_\ell} = \sum_{\alpha\in\Psi}
\sum_{\beta_1\not=\beta_2\not=\ldots\beta_\ell\in\Psi(\alpha)}
\frac{1}{(\alpha\cdot a)^n(\beta_1\cdot a)^{m_1}(\beta_2\cdot a)^{m_2}\cdots (\beta_\ell\cdot a)^{m_\ell}}~.
\end{equation}
As we will show in the following, these sums will be useful to express all higher prepotential
coefficients in a very compact way.
The properties of these sums are discussed in Appendix~\ref{secn:sums} where in particular we show that $C_{1;1}$ is identically vanishing. We therefore have
\begin{equation}
\sum_{\alpha,\beta\in\Psi} \frac{\alpha\cdot\beta}{(\alpha\cdot a)(\beta\cdot a)}
 = 4\,C_2+2\,C_{1;1}= 4\,C_2
\end{equation}
Using this in (\ref{recn2}) and integrating with respect to $E_2$, we finally obtain
\begin{equation}
\label{f2bis}
f_2 = -\frac{m^4}{24} \,E_2 \, C_2 =
-\frac{m^4}{24}\, \big(1 - 24 q - 72 q^2 -96 q^3 \cdots\big)\, C_2 ~,
\end{equation}
where in the second step we inserted the Fourier expansion of $E_2$. We observe that the $q^0$-term matches the $m^4$ contribution in the one-loop result in (\ref{fn1loop}).  The higher order terms in the $q$-expansion are a prediction for the instanton corrections to $f_2$.
As we will see in Section~\ref{snek}, these predictions can be tested and verified for the first few instanton numbers in various gauge groups of the A and D series using localization methods. For the
exceptional groups $E_{6,7,8}$, instead, these are truly predictions since the multi-instanton calculus
is not available in these cases.

We now consider the next mass order. For $f_3$, from (\ref{rec}), we have
\begin{equation}
\label{recf31}
\frac{\partial f_3}{\partial_{E_2}} = -\frac{1}{12}
\frac{\partial f_1}{\partial a}\cdot \frac{\partial f_2}{\partial a}
=  -\frac{m^6}{288}\, E_2
\sum_{\alpha,\beta\in\Psi} \frac{\alpha\cdot\beta}{(\alpha\cdot a)^3(\beta\cdot a)} ~,
\end{equation}
where we have used the explicit expressions of $f_1$ and $f_2$ given in (\ref{f1is}) and (\ref{f2bis}) to
do the second step. Manipulating the root sums as before and using the identity (\ref{C31C211}),
we can rewrite (\ref{recf31}) as
\begin{equation}
\label{recf3}
\frac{\partial f_3}{\partial_{E_2}}  =
-\frac{m^6}{72} E_2 \Big(C_4 + \frac{1}{4}C_{2;11}\Big) ~.
\end{equation}
Integrating with respect to $E_2$, we find
\begin{equation}
\label{f3is}
f_3 = - \frac{m^6}{144} E_2^2 \Big(C_4 + \frac{1}{4}C_{2;11}\Big) + x\, E_4~,
\end{equation}
where we have taken into account that the ``integration constant" must have modular weight $4$ and thus must be proportional to $E_4$. The coefficient $x$ must be chosen in such a way that in the perturbative limit, where $E_2$ and $E_4$ become 1, one recovers the $m^6$ term in the one-loop result (\ref{fn1loop}).
This requires that
\begin{equation}
\label{cis}
x = -m^6\,\Big(\frac{C_4}{720} - \frac{C_{2;11}}{576}\Big)~.
\end{equation}
Plugging this back into (\ref{f3is}), we finally obtain
\begin{equation}
\label{f3tris}
f_3 = -\frac{m^6}{720}\,\big(5 E_2^2 + E_4\big)\, C_4 -
\frac{m^6}{288}\,\big(E_2^2 - E_4\big)\times\frac{1}{2}\, C_{2;11}~.
\end{equation}
Expanding the Eisenstein series in powers of $q$ we find
\begin{equation}
f_3= -\frac{m^6}{120}\,C_4+q\,\frac{m^6}{2}\,C_{2;1,1}+q^2m^6\,\big(-6C_4+3C_{2;11}\big)
+q^3m^6\,\big(-32C_4+6C_{2;11}\big)+\cdots
\label{f3exp}
\end{equation}
from which we can explicitly read the multi-instanton corrections.

Using the recursion relation and the comparison with the perturbative expression, we have determined also the terms of order $m^8$ and $m^{10}$ in the prepotential. We now collect all our results up to $f_5$:
\begin{subequations}
\label{f5is}
\begin{align}
f_1 &= \frac{m^2}{4} \sum_{\alpha\in\Psi}  \log\left(\frac{\alpha\cdot a}{\Lambda}\right)^2~, \\
f_2 & =-\frac{m^4}{24} \,E_2 \, C_2 ~, \label{f2fin}\\
f_3 &= -\frac{m^6}{720}\,\big(5 E_2^2 + E_4\big)\, C_4 -
\frac{m^6}{288}\,\big(E_2^2 - E_4\big)\times\frac{1}{2}\, C_{2;11}~,\label{f3fin}\\
f_4 & = -\frac{m^8}{90720}\,\big(175 E_2^3 + 84 E_2 E_4 + 11 E_6\big)\, C_6 \notag\\
&~~~~+
\frac{m^8}{8640}\,\big(5 E_2^3 - 3 E_2 E_4 - 2 E_6\big) \Big(C_{4;2} +\frac{1}{12} C_{3;3}\Big)\notag\\
&~~~~-\frac{m^8}{1728}\,\big(E_2^3 - 3 E_2 E_4 + 2 E_6\big)\times\frac{1}{24} C_{2;1111}~,\label{f4fin}\\
f_5 & = -\frac{m^{10}}{362880}\,\big(245 E_2^4 + 196 E_2^2 E_4 + 44 E_2 E_6 + 19 E_4^2\big)\, C_8\notag\\
&~~~~ +\frac{m^{10}}{145152}\,\big(35 E_2^4 - 7 E_2^2 E_4 - 18 E_2 E_6 - 10 E_4^2\big)
\Big(C_{6;2} -\frac{13}{45} C_{3;3}\Big)\notag\\
& ~~~~+ \frac{m^{10}}{82944}\,\big(E_2^2-E_4\big)^2 \Big(\frac{5}{12} C_{4;4} - 3 C_{4;22} - C_{3;32} - C_{4;31}\Big)\notag\\
& ~~~~- \frac{m^{10}}{6912}\,\big(E_2^4 - 6 E_2^2 E_4  + 8 E_6 E_2 - 3 E_4^2\big)\times \frac{1}{720} C_{2;111111}~.
\end{align}
\end{subequations}
If we were to proceed to the next order, {\it{i.e.}} to $f_6$, after having determined all terms containing $E_2$, we would still have to fix a purely modular term of order 12. Since there are two independent modular forms of weight 12, namely $E_6^2$ and $E_4^3$, we could no longer
fix the coefficient of these two forms by comparison to the one-loop result only;
we would also need to know the one-instanton result, if available. Having done this, however, all the subsequent instanton corrections would be predicted. The covariance of the prepotential under S-duality,
implemented through the recursion relation, is a symmetry requirement: it is not sufficient by itself to determine the dynamics, and, in particular, it does not eliminate the need to evaluate explicitly the non-perturbative corrections. Still, it minimizes the number of such computations.

\subsection{1-instanton terms}
\label{subsec:oneinst}

Let us consider the 1-instanton terms in the prepotential. Substituting the $q$-expansion of
the Eisenstein series into (\ref{f5is}) one can see that the only terms which contribute at order $q$
are those proportional to $C_{2;11\cdots}$, whose coefficients follow an obvious pattern:
\begin{equation}
\label{f1instup3}
\begin{aligned}
F_{k=1} & =  m^4 C_2 + \frac{m^6}{2!} C_{2;11} + \frac{m^8}{4!} C_{2;1111} +
\frac{m^{10}}{6!}C_{2;111111} + \cdots \\
& =
\sum_{\alpha\in\Psi} \frac{m^4}{(\alpha\cdot a)^2}\, \sum_{\ell=0} \frac{m^{2\ell}}{\ell!}
\sum_{\beta_1\not=\beta_2\not=\ldots\beta_l\in\Psi(\alpha)} \frac{1}{(\beta_1\cdot a)\cdots (\beta_\ell\cdot a)}
\end{aligned}
\end{equation}
where in the second line we have used the explicit definition of the sums $C_{2;11\cdots}$.
This pattern extends to all orders in $m$, and we can rewrite the above expression as%
\footnote{Note that the terms with odd powers of $m$ that we obtain expanding the product in the right hand side vanish identically, as discussed in Appendix~\ref{secn:sums}.}
\begin{equation}
\label{f1qnop}
F_{k=1} =  \sum_{\alpha\in\Psi} \frac{m^4}{(\alpha\cdot a)^2}
\prod_{\beta\in\Psi(\alpha)}\left(1 + \frac{m}{\beta\cdot a}\right)~.
\end{equation}

We now show that, in the decoupling limit in which the $\cN=2^\star$ theory reduces to the
pure $\cN=2$ SYM, the above result agrees with the explicit computations that are present in the literature
\cite{Benvenuti:2010pq}-\nocite{Keller:2011ek,Hanany:2012dm}\cite{Cremonesi:2014xha}.
In the decoupling limit the mass $m$ is sent to infinity and the instanton weight $q$ to zero, keeping constant the dynamically generated scale $\widehat\Lambda$ defined as
\begin{equation}
\label{declim}
\widehat\Lambda^{\,2h^\vee} = m^{2h^\vee} \, q~.
\end{equation}
Here $2h^{\!\vee}$ is the one-loop $\beta$-function coefficient of the pure $\cN=2$ theory, expressed
in terms of the dual Coxeter number of the Lie algebra $\mathfrak{g}$. For the single-laced algebras these
numbers are given by %
\[
\begin{array}{c|ccccc}
   &  ~~A_r~~ &  ~~  D_r~~  & ~~E_6~~ &~~E_7 ~~& ~~E_8~~  \\
\hline
h^\vee \phantom{\Big|}  &   r  & 2r-2  & 12 & 18 & 30     \\
\end{array}
\]
Since the number of roots $\beta$ in the set $\Psi(\alpha)$ is $2h^\vee-4$, the highest mass power in (\ref{f1qnop}) is exactly $m^{2h^\vee}$, and so it is consistent to take the decoupling limit,
in which all terms proportional to non-maximal powers of $m$ vanish. Doing this, we remain with
\begin{equation}
\label{purek1}
q\,F_{k=1}
~\longrightarrow~
\widehat\Lambda^{\,2h^\vee} \sum_{\alpha\in\Psi} \frac{1}{(\alpha\cdot a)^2}
\prod_{\beta\in\Psi(\alpha)} \frac{1}{\beta\cdot a}~.
\end{equation}
This expression has been derived in \cite{Keller:2011ek} following a completely different approach%
\footnote{In \cite{Keller:2011ek} also the non-simply laced groups are considered; in the companion paper
\cite{Billo} we will show that also in these cases our treatment reproduces, in the decoupling limit, their expression.}. Our result in (\ref{f1qnop})
generalizes this to the $\cN=2^\star$ case.

\section{The recursion relation in the $\Omega$-background}
\label{sec:recepsi}

The general features described in the previous section hold also when the $\cN=2^\star$
theories are formulated in an $\Omega$-background \cite{Nekrasov:2003rj}. In fact we are
going to show that for a generic gauge group of the ADE series the $\Omega$-deformed prepotential
satisfies a generalized recursion relation, thus extending the analysis of \cite{Huang:2006si}\,-\,\nocite{Grimm:2007tm,Huang:2009md,Huang:2010kf,Huang:2011qx,Galakhov:2012gw,Billo:2013fi,Billo:2013jba}\cite{Nemkov:2013qma}
for the SU$(2)$ theory and of \cite{Billo:2014bja} where the SU$(N)$ theories were considered.

We parametrize the $\Omega$-background by $\epsilon_1$ and $\epsilon_2$ and, for later
convenience, introduce the following combinations
\begin{equation}
\epsilon=\epsilon_1+\epsilon_2~,\qquad h=\sqrt{\epsilon_1\epsilon_2}~.
\label{epsilonh}
\end{equation}
The deformed prepotential can still be written as in (\ref{iprep1})-(\ref{fn1linst}), but both
the one-loop and the instanton parts receive corrections in $\epsilon$ and $h$.
In particular, the one-loop term becomes \cite{Nekrasov:2003rj,Huang:2011qx,Billo:2013fi}
\begin{equation}
\begin{aligned}
f^{\mathrm{1-loop}} &= h^2 \sum_{\alpha\in\Psi}
\Big[\log\Gamma_2(\alpha\cdot\phi)-\log\Gamma_2(\alpha\cdot\phi+m+\epsilon)\Big]
\end{aligned}
\label{F1loopep}
\end{equation}
where $\Gamma_2$ is the Barnes double $\Gamma$-function (see Appendix~\ref{secn:appgamma2}).
By expanding for small values of $m$, $\epsilon$ and $h$, we obtain
\begin{subequations}
\label{f1loopep}
\begin{align}
f_1^{\mathrm{1-loop}} &= \frac{M^2}{4} \sum_{\alpha\in\Psi}
\log\Big(\frac{\alpha\cdot\phi}{\Lambda}\Big)^2~, \\
f_2^{\mathrm{1-loop}} &= - \frac{M^2(M^2 +h^2)}{24}\, C_2 ~,
\label{f2he}\\
f_3^{\mathrm{1-loop}} &= - \frac{M^2(M^2 +h^2)(2M^2+ 3h^2 -\epsilon^2)}{240}\, C_4 ~,
\label{f3he}\\
f_4^{\mathrm{1-loop}} &= - \frac{M^2(M^2 +h^2)
(3M^4 + 10 h^4 +2 \epsilon^4 + 11h^2M^2 -4\epsilon^2M^2 -10 h^2\epsilon^2)}{1008}\, C_6 ~,
\label{f4he}
\end{align}
\end{subequations}
where we have defined
\begin{equation}
 M^2 \,\equiv\, m^2- \frac{\epsilon^2}{4}~.
 \label{M2}
\end{equation}
As in the undeformed case, also here $f_1$ does not receive instantonic corrections, so we have
\begin{equation}
f_1=f_1^{\mathrm{1-loop}}~,
\end{equation}
while all other $f_n$'s with $n\geq 2$ have contributions at any order in the instanton expansion.

The exact $q$-dependence of the deformed $f_n$'s can be determined by requiring that
the prepotential transforms properly under S-duality. In an $\Omega$-background
this means that S-duality acts on the prepotential as a Fourier
transform \cite{Galakhov:2012gw}\nocite{Billo:2013fi,Billo:2013jba}\,-\,\cite{Nemkov:2013qma}, namely
\begin{equation}
\exp\Big(\!-\frac{S[F](a^{\mathrm{D}})}{h^2}\Big)=
\Big(\frac{\ii \tau}{h^2}\Big)^{r/2}
\int d^{\,r} x ~\exp\Big(\frac{2\pi \ii \,a^{\mathrm{D}} \cdot x - F(x)}{h^2}\Big)
\label{FT2}
\end{equation}
where $r$ is the rank of the gauge group. 
This interpretation of S-duality is fully consistent with the
interpretation of $a$ and $a^{\mathrm{D}}$ as canonical conjugate variables, on which
S acts as a canonical transformation, and of
\begin{equation}
Z= \exp\Big(\!-\frac{F}{h^2}\Big)
\end{equation}
as a wave function in a quantization of this phase space, with $h^2 = \eud$ playing the r\^ole of the Planck
constant \cite{Witten:1993ed}\nocite{Aganagic:2006wq}\,-\,\cite{Gunaydin:2006bz}.

If we compute the Fourier transform (\ref{FT2}) in the saddle point approximation for $h \to 0$ and 
denote by $a$ the solution of the saddle
point equation, that is
\begin{equation}
 2\pi \ii \,a^{\mathrm{D}}- \partial_x F(x)\Big|_{x=a}=0~,
\end{equation}
then the leading contribution to the integral in (\ref{FT2}) is
\begin{equation}
\exp\Big(\!-\frac{S[F](a^{\mathrm{D}})}{h^2}\Big)=
\exp\left(\!
-\frac{F(a) - a\cdot\partial_a F(a)}{h^2}
-\frac12 \,\mathrm{tr} \log\Big(\delta_{u v}+\frac{1}{2\pi \ii \tau} \, \partial_u \partial_v f \Big) \right) +\cdots
\label{FT3}
\end{equation}
where the tr log part comes from the Gaussian integration around the saddle point and the ellipses stand
for subleading terms in $h$. The dominant contribution for $h\to0$ reproduces
the Legendre transform of the prepotential as expected, but there are corrections for finite $h$.
Indeed we have
\begin{equation}
\begin{aligned}
S[F]&=\cL[F]+\frac{h^2}{2}\,
\mathrm{tr} \log\Big(\delta_{u v}+\frac{1}{2\pi \ii \tau} \, \partial_u \partial_v f \Big) +\cdots\\
&=\cL[F]+\delta\,\Big(\frac{h^2}{24}\Delta f\Big)+\cO(\delta^2)+\cdots
\end{aligned}
\label{FT4}
\end{equation}
where we have used $\delta=\ft{6}{\pi\ii\tau}$, as before, and defined $\Delta=\sum_u\partial_u^2$.

Repeating the same steps described in Section~\ref{secn:sduality} (see also Sections 3 and 4 of \cite{Billo:2013jba} for more details), one can show that (\ref {FT4}) leads to the
following recursion relation for the prepotential coefficients $f_n$'s:
\begin{equation}
\frac{\partial f_n}{\partial E_2}=-\frac{1}{24}\sum_{m=1}^{n-1} \frac{\partial f_m}{\partial a}\cdot
\frac{\partial f_{n-m}}{\partial a}\,+\,\frac{h^2}{24}\,\Delta f_{n-1}~.
\label{recursion}
\end{equation}
The recursive computation of the $f_n$'s can then proceed along the lines we have discussed
in the undeformed theory. At level two we find
\begin{equation}
\frac{\partial f_2}{\partial E_2}=-\frac{1}{24}\frac{\partial f_1}{\partial a}\cdot
\frac{\partial f_1}{\partial a}\,+\,\frac{h^2}{24}\,\Delta f_1= -\frac{1}{24} M^2 (M^2+h^2) \,C_2
\label{rec-f2}
\end{equation}
where we have used
\begin{equation}
\Delta f_1 = -\frac{M^2}{2} \sum_{\alpha\in\Psi}
\frac{(\alpha \cdot\alpha)}{(\alpha \cdot\phi)^2} = -M^2\, C_2~.
\end{equation}
Integrating with respect to $E_2$ we get
\begin{equation}
f_2=-\frac{1}{24} M^2 (M^2+h^2)\, E_2\, C_2~.
\label{f2ep}
\end{equation}
It is immediate to see that in the perturbative limit, when $E_2$ reduces to 1, this correctly reproduces
(\ref{f2he}).

Using this result we can write the differential equation constraining $f_3$, namely
\begin{equation}
\begin{aligned}
\frac{\partial f_3}{\partial E_2} &=
-\frac{1}{12}\frac{\partial f_1}{\partial a}\cdot
\frac{\partial f_2}{\partial a}\,+\,\frac{h^2}{24}\,\Delta f_2 \\
&= -\frac{1}{144} M^2 (M^2+h^2)
\big((2M^2+3h^2) C_4 + M^2 C_{3;1} \big)\,E_2~.
\end{aligned}
\label{rec-f3}
\end{equation}
Integrating with respect to $E_2$ and fixing the dependence on $E_4$ in such a way to
reproduce the perturbative result (\ref{f3he}), we get
\begin{equation}
\begin{aligned}
f_3&=-\frac{1}{288} M^2 (M^2+h^2)\Big[\frac{1}{5}\big((2M^2+3h^2)(5 E_2^2+E_4)-6\epsilon^2E_4\big)\,C_4
\\&~~~\qquad
+\frac{1}{2} M^2\,(E_2^2-E_4)\,C_{2;11}\Big]
\end{aligned}
\label{f3ep}
\end{equation}
where we have used the identity $C_{3;1}=C_{2;11}$ proven in Appendix \ref{secn:sums}.

In a similar way we can determine $f_4$. The result we get is
\begin{eqnarray}
 f_4 &=&-\frac{1}{1728} M^2 (M^2+h^2) \left\{
\Big[
\frac{2}{105}(2M^2+3h^2)(2M^2+5h^2)(35 E_2^3+21E_2E_4+4E_6)
\right.\nonumber\\
&&
+\frac{2}{21} M^2(M^2+h^2)(7 E_2^3-E_6)
-\frac{12}{35} \epsilon^2 (2M^2+5h^2)(7E_2E_4+3E_6)
+\frac{24}{7} \epsilon^4E_6 \Big]C_6 \nonumber\\
&&
-\Big[
\frac{1}{10}M^2(2M^2+3h^2)(5 E_2^3-3E_2E_4-2E_6)-
\frac{6}{5} \epsilon^2 M^2(E_2E_4-E_6) \Big] C_{4;2} \nonumber \\
&&
-\Big[
\frac{1}{60}M^2(M^2+4h^2)(5 E_2^3-3E_2E_4-2E_6) -
\frac{3}{5} \epsilon^2 M^2 (E_2E_4-E_6) \Big] C_{3;3} \nonumber \\
&&\left.
+\frac{1}{24} M^4 (E_2^3-3E_2E_4+2E_6)\, C_{2;1111}\right\}~.
\label{f4ep}
\end{eqnarray}
This procedure can be carried out for the next orders in the mass expansion, but the results
become lengthy and we see no reason to explicitly report them.

\subsection{1-instanton terms}
While the exact expressions of the $f_n$'s are rather involved, their 1-instanton part is quite simple
and it is possible to write a compact expression which generalizes the one we have derived in
(\ref{f1qnop}) for the underformed theory. Indeed, inserting the $q$-expansions of the Eisenstein series
in (\ref{f2ep}), (\ref{f3ep}) and (\ref{f4ep}) one can see that only few terms contribute at order $q$:
\begin{equation}
\label{fk1}
\begin{aligned}
f_2\big|_{k=1}&=M^2(M^2+h^2)\, C_2 ~,\\
f_3\big|_{k=1}&=M^2(M^2+h^2)\,\Big[\epsilon^2 C_4 + \frac{1}{2} M^2C_{2;11} \Big] ~,\\
f_4\big|_{k=1}&=M^2(M^2+h^2)\, \Big[\epsilon^4 C_6 + \frac{1}{2} \epsilon^2M^2C_{4;11}+\frac{1}{24} M^2(M^2-
\epsilon^2 ) C_{2;1111}\Big]
\end{aligned}
\end{equation}
where in the last equation we have used the first identity given in (\ref{identfin}). Actually this pattern extends also to higher $f_n$'s, as we have verified by computing the 1-instanton prepotential using localization techniques described in the next section. Altogether we find
\begin{equation}
\label{FFk1}
\begin{aligned}
F_{k=1}&=\sum_{n=2} f_n\big|_{k=1} \\
&= M^2(M^2+h^2)\,
\Big[ \big(C_2 +\epsilon^2\, C_4 + \epsilon^4\, C_6 + \epsilon^6 \,C_8\cdots\big) \\
&\qquad\quad + \frac{1}{2} M^2 \big(C_{2;11}+ \epsilon^2\, C_{4;11}+ \epsilon^4 \,C_{6;11} + \cdots\big)
\\
&\qquad\quad +\frac{1}{24} M^2(M^2-\epsilon^2 ) \big(C_{2;1111}+ \epsilon^2\, C_{4;1111}
+ \cdots \big)\\
&\qquad\quad+\frac{1}{720} M^2(M^4-3M^2\epsilon^2+3\epsilon^4) \big(C_{2;111111}+ \cdots \big)
+\cdots\Big]~.
\end{aligned}
\end{equation}
This pattern suggests to introduce the following notation
\begin{equation}
\begin{aligned}
g_{2n}&=\frac{1}{(2n)!}\Big(
C_{2;{\underbrace{\mbox{\scriptsize{11\ldots1}}}_{\mbox{\scriptsize{$2n$}}}}}+
\epsilon^2\, C_{4;{\underbrace{\mbox{\scriptsize{11\ldots1}}}_{\mbox{\scriptsize{$2n$}}}}}+
\epsilon^4\,C_{6;{\underbrace{\mbox{\scriptsize{11\ldots1}}}_{\mbox{\scriptsize{$2n$}}}}}
+\cdots\Big)\\
&=
\frac{1}{(2n)!}\,\sum_{\alpha\in \Psi} \sum_{\beta_1 \neq \cdots \beta_{2n} \in \Psi(\alpha)}
\frac{1}{(\alpha \cdot a)(\alpha \cdot a+\epsilon)(\beta_1 \cdot a)
\cdots(\beta_{2n} \cdot a) }~,
\end{aligned}
\label{g2n}
\end{equation}
so that (\ref{FFk1}) becomes
\begin{equation}
F_{k=1}= M^2(M^2+h^2)\Big[g_0+M^2\,g_2+M^2(M^2-\epsilon^2)\,g_4+
M^2(M^4-3M^2\epsilon^2+3\epsilon^4) \,g_6+\cdots\Big]~.
\label{FFk1a}
\end{equation}
Notice that in the sums $g_{2n}$ all odd powers in the $\epsilon$-expansion of the second line
of (\ref{g2n}) vanish upon summing over the roots and
that, for a given algebra $\mathfrak{g}$, the highest non-vanishing expression of this kind is
$g_{2h^{\!\vee}-4}$, since the order of $\Psi(\alpha)$ is $2h^{\!\vee}-4$.
It is interesting to observe that the $g$'s can be expressed in terms of a generating function
\begin{equation}
\label{GG}
G(x)= \sum_{n=0}^{2h^{\!\vee}-4}g_n\,x^n=
\sum_{\alpha\in \Psi}\frac{1}{(\alpha \cdot a)(\alpha \cdot a+\epsilon)}
\prod_{\beta \in \Psi(\alpha)}\left( 1+\frac{x}{\beta \cdot a}\right)
\end{equation}
where
\begin{equation}
\label{gniader}
g_{n} = \frac{1}{n!} \left.\frac{\partial ^n G(x)}{\partial x^n}\right|_{x=0}~.
\end{equation}
It is also possible to recognize a pattern in the mass- and $\epsilon$-dependent expressions that multiply the $g_n$'s in \eq{FFk1a}. Writing the latter in the form
\begin{equation}
\label{F2Fk1}
F_{k=1}= M^2(M^2+h^2)\,\sum_{n=0}^{2h^\vee-4} g_n \,\epsilon^{n} H_n\Big(\ft{M^2}{\epsilon^2}\Big)~,
\end{equation}
one can see that the polynomials $H_n$ are connected to the Euler polynomials $\cE_n$ according to
\begin{equation}
\label{H2}
H_n\Big(\ft{M^2}{\epsilon^2}\Big)= \frac{1}{2} \left[ \cE_n\Big( \ft{1}{2}+\ft{m}{\epsilon} \Big) +
\cE_n\Big( \ft{1}{2}-\ft{m}{\epsilon}\Big)  \right]
\end{equation}
(recall that $M^2 = m^2-\ft{\epsilon^2}{4}$).
In turn, the Euler polynomials are defined  by
\begin{equation}
\label{defeulerpol}
\frac{2\,\ee^{z\,t}}{\ee^t + 1} = \sum_{n=0}^\infty\frac{1}{n!}\, \cE_n(z)\,t^n~.
\end{equation}
With this definition one can easily check that all $H_{2n+1}$ are vanishing, while the $H_{2n}$ reproduce the
expressions appearing in (\ref{FFk1a}). Inserting (\ref{H2}) and (\ref{gniader}) into (\ref{F2Fk1}), we then
obtain
\begin{equation}
\label{F3Fk1}
\begin{aligned}
F_{k=1} & = \frac 12\,\Big(m^2-\frac{\epsilon^2}{4}\Big)\Big(m^2-\frac{\epsilon^2}{4}+h^2\Big)\\
&~~~\times
\sum_{n=0}^\infty  \frac{1}{n!}\, \left[ \cE_n\Big( \ft{1}{2}+\ft{m}{\epsilon} \Big) +
\cE_n\Big( \ft{1}{2}-\ft{m}{\epsilon}\Big)  \right]
\left(\epsilon \,\frac{\partial}{\partial x}\right)^n \!G(x)\Big|_{x=0}~.
\end{aligned}
\end{equation}
Since $G(x)$ is a polynomial of order $2h^{\!\vee} -4$, all terms with $n>2h^{\!\vee} -4$ in the sum vanish
and (\ref{F3Fk1}) is simply another way to write (\ref{F2Fk1}). 
However, this allows us to use the property (\ref{defeulerpol}) of the Euler polynomials
in order to write
\begin{eqnarray}
F_{k=1} & =& \Big(m^2-\frac{\epsilon^2}{4}\Big)\Big(m^2-\frac{\epsilon^2}{4}+h^2\Big)\,
\Bigg(\,\frac{\ee^{(\frac{\epsilon}{2}+m)\,\partial_x}}{\ee^{\epsilon\,\partial_x}+1}+
\frac{\ee^{(\frac{\epsilon}{2}-m)\,\partial_x}}{\ee^{\epsilon\,\partial_x}+1}
\,\Bigg) G(x)\Big|_{x=0}\label{F3Fk1a}\\
& =& \Big(m^2-\frac{\epsilon^2}{4}\Big)\Big(m^2-\frac{\epsilon^2}{4}+h^2\Big)\!
\sum_{n=0}^{2h^\vee-4}\!\left(-\epsilon\,\frac{\partial}{\partial x}\right)^{\!n}
\Big[G\big(x + \ft{\epsilon}{2} + m\big) + G\big(x + \ft{\epsilon}{2} - m\big)\Big]\Big|_{x=0}
\nonumber
\end{eqnarray}
where in the second line we truncated the expansion of the geometric series since, as we stressed above,
$G(x)$ is a polynomial of order $2h^{\!\vee} -4$. It is not difficult to check that
in the limit $\epsilon \to 0$ we recover the 1-instanton expression given in (\ref{f1qnop}) and that
keeping the $\epsilon$ but decoupling the matter hypermultiplet we recover the same formula obtained
in \cite{Keller:2011ek} for the pure $\cN=2$ theories from the coherent states of the W-algebras.

\section{Multi-instanton calculations}
\label{snek}

In this section, we test the results for the ${\mathcal N}=2^\star$ prepotential obtained from the modular recursion equation against a direct microscopic computation of the first instanton corrections
based on equivariant localization techniques \cite{Nekrasov:2002qd}\nocite{Flume:2002az,Nekrasov:2003rj,Bruzzo:2002xf}\,-\,\cite{Fucito:2004ry} (see also \cite{Billo:2012st} for further details).
To do so we first recall a few basic facts about the instanton moduli space and the multi instanton
calculus starting from the gauge theories with unitary groups.

\subsection{Multi-instantons for the U$(N)$ gauge theory}

The moduli space of $k$-instantons in the $\mathcal{N}=2^\star$ theory with gauge group U$(N)$ can
be built from the open strings connecting a stack of $k$ D(-1) and $N$ D3-branes in Type IIB string theory.
The gauge theory prepotential can be viewed as the free energy of the statistical system describing the
lowest modes of the open strings with at least one end-point on the D(-1) branes that account for the
the instanton moduli \cite{Witten:1995gx}\nocite{Douglas:1995bn,Green:2000ke,Billo:2002hm}\,-\,\cite{Billo:2006jm}.

The partition function $Z_k$ of the system can be computed using localization methods. To achieve full localization all symmetries of the system have to be broken. The gauge symmetries on the world volumes of the
D3 and D(-1) branes can be broken by distributing them along a transverse complex plane $\mathbb{C}$.
We label their positions in this plane by $a_u$ (with $u=1,\cdots, N$) and $\chi_i$ (with $i=1,\cdots,k$), respectively.
The  $\mathrm{SO}(4)\times \mathrm{SO}(4)$ Lorentz symmetry of the spacetime transverse to this plane can be broken  by turning on an $\Omega$-background with parameters  $\epsilon_1$, $\epsilon_2$, $\epsilon_3$ and $\epsilon_4$.
The first two parameters, $\epsilon_1$ and $\epsilon_2$, describe a gravitational background, while
$\epsilon_3$ and $\epsilon_4$ are related to the mass of the adjoint hypermultiplet.

In Tab.~1 we list all moduli for given $k$ and $N$, together with
their transformation properties with respect to the various symmetry groups.
In the first column we have grouped the moduli into $Q$-pairs of the supersymmetric charge $Q$ used for localization and labeled by their $\mathrm{SO}(4)\times \mathrm{SO}(4)$ quantum numbers
with spinor indices $\alpha,\dot\alpha,a,\dot a$, all taking two values.
\begin{table}[ht]
\begin{center}
\begin{tabular}{|c|c|c|c|}
\hline
\begin{small} $ (\phi,\psi) $ \end{small}
&\begin{small} $(-1)^{F_\phi}$ \end{small}
&\begin{small} $\mathrm{U}(N) \times \mathrm{U}(k) $
\end{small}
&\begin{small}  $ \lambda_\phi\phantom{\Big|} $ \end{small}
\\
\hline\hline
$(B_{\alpha\dot\alpha},M_{\alpha \dot a} )$ & $+\phantom{\Big|}$ & $\bigl(\bone,\Yfund\,\overline{\Yfund}\bigr)$
&   $ \chi_{ij}+\epsilon_1,\,\chi_{ij}+\epsilon_2$\\
$(B_{ a \dot a},M_{\dot\alpha  a} )$ & $+\phantom{\Big|}$  & $\bigl(\bone,\Yfund\,\overline{\Yfund}\bigr)$
&   $ \chi_{ij}+\epsilon_3,\chi_{ij}+\epsilon_4$\\
$(N_{(\dot\alpha\dot b)},D_{(\dot\alpha\dot \beta)})$ & $-\phantom{\Big|}$
& $\bigl(\bone,\Yfund\,\overline{\Yfund}\bigr)  $ & $ \sqrt{\chi_{ij}},\, \chi_{ij}+\epsilon_1+\epsilon_2  $\\
$(\bar \chi , N)$ & $+\phantom{\Big|}$
& $\bigl(\bone,\Yfund\,\overline{\Yfund}\bigr)  $ & $ \sqrt{\chi_{ij}}   $\\
$(N_{\alpha a},D_{\alpha a})$ & $-\phantom{\Big|}$
& $\bigl(\bone,\Yfund\,\overline{\Yfund}\bigr)  $ & $ \chi_{ij}+\epsilon_1+\epsilon_3, \,\chi_{ij}+\epsilon_1+\epsilon_4  $\\
$(w_{\dot \alpha},\mu_{\dot a})$ & $+\phantom{\Big|}$
& $\bigl( \overline{\Yfund},\Yfund \bigr)  $
   & $ \chi_i-a_u+\ft{\epsilon_1+\epsilon_2}{2}    $  \\
  $(\bar w_{\dot \alpha},\bar \mu_{\dot a})$ & $+\phantom{\Big|}$
& $\bigl({\Yfund},\overline{\Yfund}\bigr)  $
   & $ -\chi_i+a_u+\ft{\epsilon_1+\epsilon_2}{2}    $  \\
     $(h_{a},\mu_{a})$ & $-\phantom{\Big|}$
& $\bigl(\overline{\Yfund}, \Yfund \bigr)  $
   &   $ \chi_i-a_u+\ft{\epsilon_3-\epsilon_4 }{2}  $  \\
  $(\bar h_{a},\bar \mu_{a})$ & $-\phantom{\Big|}$
& $\bigl({\Yfund} , \overline{\Yfund}\bigr)  $
   &   $ -\chi_i+a_u+\ft{\epsilon_3-\epsilon_4 }{2}  $  \\
\hline
\end{tabular}
\caption{Instanton moduli for the U$(N)$ gauge theory. The columns display, respectively, the moduli in a
ADHM-like notation, their statistics,
their transformation properties with respect to the gauge and instanton symmetry groups and their
$Q^2$-eigenvalues $\lambda_\phi$. The notation $\chi_{ij}$ stands for $\chi_i-\chi_j$.}
\label{tabun}
\end{center}
\end{table}
The neutral bosonic moduli include the eight instanton positions transverse to $\C$, denoted by $B_{\alpha\dot\alpha}$ and $B_{a\dot a}$, and the positions along $\C$, denoted by $\chi$ and $\bar\chi$. The charged bosonic moduli $w_{\dot\alpha}$ and $\bar w_{\dot\alpha}$ describe the size and the orientation of the instantons, while the auxiliary fields
$D_{\dot\alpha\dot\beta}$, $D_{\alpha a}$ and $h_a$ take care of the generalised ADHM constraints.
The field $\chi$ can be viewed as the U$(k)$ gauge parameter and thus it should be integrated out in order
to achieve U$(k)$-invariance.

The $k$-instanton partition function is given by the complex superdeterminant of $Q^2$, which can be computed from the data reported in the last column of the above table. The result is
\begin{equation}
Z_k=  \oint\prod_{i=1}^k\frac{d\chi_i}{2\pi\ii}~ \Delta(0)\,  \prod_\phi \lambda_\phi^{(-1)^{F_\phi+1} }
\,=\,
\oint\prod_{i=1}^k\frac{d\chi_i}{2\pi\ii}~z_k^{\mathrm{gauge}}\,z_k^{\mathrm{matter}}
\label{Zkun}
\end{equation}
where $\Delta(0)=\prod_{i\neq j} \chi_{ij}$ is the Vandermonde determinant and
\begin{subequations}
\begin{align}
z_k^{\mathrm{gauge}}&=\frac{(-1)^k}{k!}\,  \left( \frac{\epsilon_1+\epsilon_2 }{\epsilon_1\epsilon_2}\right)^k \,\frac{\Delta(0)\,\Delta(\epsilon_1+\epsilon_2)}{\Delta(\epsilon_1)\,\Delta(\epsilon_2)}\,\prod_{i=1}^k
\frac{ 1}{P\big(\chi_i+\frac{\epsilon_1+\epsilon_2}{2}\big)\,P\big(\chi_i-\frac{\epsilon_1+\epsilon_2 }{2}\big)}
~,\\
\notag\\
z_k^{\mathrm{matter}}&=\left( \frac{ (\epsilon_1+\epsilon_3)    (\epsilon_1+\epsilon_4)   }{   \epsilon_3 \epsilon_4 }\right)^{\!k}
\frac{\Delta(\epsilon_1+\epsilon_3)\,\Delta(\epsilon_1+\epsilon_4)}{\Delta(\epsilon_3)\,\Delta(\epsilon_4)}\,
\prod_{i=1}^k P\big(\chi_i+\ft{\epsilon_3-\epsilon_4}{2}\big)\,P\big(\chi_i-\ft{\epsilon_3-\epsilon_4}{2} \big)
\end{align}
\end{subequations}
with
\begin{equation}
\begin{aligned}
P(x) =\prod_{u=1}^N\big(x- a_u)~,\qquad
\Delta(x)=\prod_{i<j}^k\big( x^2- \chi_{ij}^2 \big) ~.
\end{aligned}
\end{equation}
The integrals in (\ref{Zkun}) are computed by closing the contours in the upper-half complex
$\chi_i$-planes after giving $\epsilon_1$, $\epsilon_2$,
$\epsilon_3$ and $\epsilon_4$ an imaginary part with the following prescription
\begin{equation}
\mathrm{Im}(\epsilon_4)\gg \mathrm{Im}(\epsilon_3)\gg\mathrm{Im}(\epsilon_2)\gg\mathrm{Im}(\epsilon_1)> 0.
\label{prescription}
\end{equation}
This choice allows us to unambiguously compute all integrals in (\ref{Zkun}) and obtain the
instanton partition of the U$(N)$ theory
\begin{equation}
Z_{\mathrm{inst}}=1+\sum_{k=1}q^k\,Z_k
\end{equation}
where $q=\ee^{2\pi\ii\tau}$. At the end of the calculations we have to set
\begin{equation}
\epsilon_3=m-\frac{\epsilon_1+\epsilon_2}{2}~,\quad
\epsilon_4=-m-\frac{\epsilon_1+\epsilon_2}{2}   \label{mass34}
\end{equation}
in order to express the result in terms of the
hypermultiplet mass $m$ in the normalization of the previous sections. The prepotential is then given by
\begin{equation}
F_{\mathrm{inst}}=-\epsilon_1\epsilon_2\,\log Z_{\mathrm{inst}}=\sum_{k=1}q^k\,F_k~;
\label{FZ}
\end{equation}
by taking the limit $\epsilon_1,\epsilon_2\to0$ one finally recovers the prepotential of the undeformed
gauge theory.

\subsubsection*{1-instanton terms}

At $k=1$ there is one integral to compute; it is very easy to see that the poles of
(\ref{Zkun}) are located at
\begin{equation}
\chi_1=a_u+\frac{\epsilon_1+\epsilon_2}{2}~.
\end{equation}
Evaluating the residues, using (\ref{mass34}) and summing over $u$ we find
\begin{equation}
F_{k=1}=-\epsilon_1 \epsilon_2\, Z_1=-\left(m^2-\frac{(\epsilon_1-\epsilon_2)^2}{4} \right)
\sum_{u=1}^N  \prod_{v\neq u}  \frac{(a_{uv }+
\ft{\epsilon_1+\epsilon_2}{2})^2-m^2}{a_{uv }(a_{uv}+\epsilon_1+\epsilon_2)}
\label{Fk1nek}
\end{equation}
where $a_{uv}=a_u-a_v$.
For example for U(2) we have
\begin{equation}
F_{k=1}\Big|_{\mathrm{U}(2)}=(M^2+h^2)\Big[-2+\frac{M^2}{a_{12}(a_{12}+\epsilon)}+
\frac{M^2}{a_{21}(a_{21}+\epsilon)}\Big]
\end{equation}
where $M^2$ and $\epsilon$ are defined in (\ref{M2}) and (\ref{epsilonh}). Notice that
the terms proportional to $M^2$ in the square brackets precisely reconstruct the sum $g_0$ defined
in (\ref{g2n}).
To get the prepotential for the SU(2) theory we simply have to set $a_1=-a_2=a$ in the above
expression; in this way we get
\begin{equation}
F_{k=1}\Big|_{\mathrm{SU}(2)}=\frac{2(M^2+h^2)\,(M^2+\epsilon^2-4 a^2)}{4 a^2-\epsilon^2}~.
\label{Fk1SU2}
\end{equation}
For unitary groups of higher rank, the expanded expression of the 1-instanton prepotential
is more cumbersome; however it is possible to check that
(\ref{Fk1nek}) can be written as%
\footnote{We discard all $a$-independent terms.}
\begin{equation}
F_{k=1}= M^2(M^2+h^2)\Big[g_0+M^2\,g_2+M^2(M^2-\epsilon^2)\,g_4+
M^2(M^4-3M^2\epsilon^2+3\epsilon^4) \,g_6+\cdots\Big]
\label{Fk1gn}
\end{equation}
in agreement with (\ref{FFk1a}). The equality between (\ref{Fk1nek}) and (\ref{Fk1gn}) (up to
$a$-independent terms) is not immediate to see, but nevertheless it is true. In the undeformed
theory, {\it{i.e.}} $\epsilon,h\to0$ and $M^2\to m^2$ the above results further simplify and reduce to
those obtained long ago in \cite{Ennes:1999fb} using a completely different approach.

 \subsubsection*{2-instanton terms}
 At $k=2$ there are two integrals in (\ref{Zkun}) to compute. The procedure we have described above is
 straightforward to implement and with the prescription (\ref{prescription}) no ambiguities arise. To
 avoid long formulas we write some explicit 2-instanton terms only in the $\epsilon,h\to 0$ limit.
 For example in the undeformed U(2) theory, we get
 \begin{equation}
 \begin{aligned}
 F_{k=2}\Big|_{\mathrm{U}(2)}&= -3 m^2+6 m^4\,\frac{1}{a_{12}^2}-12m^6\,\frac{1}{a_{12}^4}
 +5m^8\,\frac{1}{a_{12}^6}
 +\cdots\\
 &=-3m^2+3m^4\, C_{2}-6m^6\,C_{4}+\frac{5m^8}{2}\,C_{6}+\cdots
 \end{aligned}
 \end{equation}
 where in the second line we have used the sums $C_n$'s defined in (\ref{defCn}) and the dots
 stand for subleading terms in the mass expansion.

 Likewise for U(3) we find
 \begin{equation}
 \begin{aligned}
 F_{k=2}\Big|_{\mathrm{U}(3)}&= -\frac{9m^2}{2}+6m^4\Big(\frac{1}{a_{12}^2}+\frac{1}{a_{13}^2}
 +\frac{1}{a_{23}^2}\Big)\\
 &~-12m^6\Big(\frac{1}{a_{12}^4}+\frac{1}{a_{13}^4}
 +\frac{1}{a_{23}^4}+\frac{a_1^2+a_2^2+a_3^2-a_1a_2-a_1a_3-a_2a_3}{a_{12}^2a_{13}^2a_{23}^2}\Big) +\cdots\\
 &=-\frac{9m^2}{2}+3m^4\, C_{2}-6m^6\,C_{4}+3m^6\,C_{2;11}+\cdots
 \end{aligned}
 \end{equation}
 We have explicitly checked up to U(5) that the same pattern occurs, namely that the 2-instanton prepotential
 is (up to $a$-independent terms)
 \begin{equation}
 F_{k=2}=3m^4\, C_{2}-6m^6\,C_{4}+3m^6\,C_{2;11}+\frac{5m^8}{2}\,C_6+6m^8\,C_{4;2}+\frac{m^8}{2}\,C_{3;3}+\frac{m^8}{2}\,C_{2;1111}+\cdots~.
 \label{Fk2fin}
 \end{equation}
This result is in total agreement with the 2-instanton prepotential that can be obtained from (\ref{f5is}) by expanding the Eisenstein series; moreover it clearly shows the advantage of using the root lattice sums $C_{n;m_1\cdots}$ that allow us to write a single expression valid for all U$(N)$'s groups.

Finally we mention that for the unitary groups it is possible to push the calculations to higher instanton numbers
as we have shown in \cite{Billo:2014bja}.

 \subsection{Multi-instantons for the SO$(2N)$ gauge theory}

The moduli space of the SO$(2N)$ gauge theory is obtained from that of the U$(2N)$ theory by using
the projector $(1+\Omega\, I)/2$  where $\Omega$ is the orientifold operator that
changes the orientation of the open strings and $I$ reflects the moduli carrying an index $\dot\alpha$ ,
{\it {i.e.}} transforming in the fundamental representation of the $SU(2)_{\mathrm{L}}$ factor of the
spacetime Lorentz group \cite{Fucito:2004gi}.
As a result, the symmetry of the brane system reduces to $\mathrm{SO}(2N)\times \mathrm{Sp}(2k)$.
 The instanton moduli and their transformation properties are listed in Tab.~2 which uses the same notation
 as Tab.~1.
\begin{table}[ht]
\begin{center}
\begin{tabular}{|c|c|c|c|}
\hline
\begin{small} $ (\phi,\psi) $ \end{small}
&\begin{small} $(-1)^{F_\phi}$ \end{small}
&\begin{small} $\mathrm{SO}(2N) \times \mathrm{Sp}(2k)$
\end{small}
&\begin{small}  $ \lambda_\phi\phantom{\Big|} $ \end{small}
\\
\hline\hline
$(B_{\alpha\dot\alpha},M_{\alpha \dot a} )$ & $+\phantom{\Big|}$ & $\bigl(\bone,\Yasymm\bigr)$
&   $ \chi_{ij}+\epsilon_1,\,\chi_{ij}+\epsilon_2$\\
$(B_{ a \dot a},M_{\dot\alpha  a} )$ & $+\phantom{\Big|}$ & $\bigl(\bone,\Ysymm\bigr)$
&   $ \chi_{ij}+\epsilon_3,\,\chi_{ij}+\epsilon_4$\\
$(N_{(\dot\alpha\dot b)},D_{(\dot\alpha\dot \beta)})$ & $-\phantom{\Big|}$
& $\bigl(\bone,\Ysymm\bigr)  $ & $ \sqrt{\chi_{ij}},\, \chi_{ij}+\epsilon_1+\epsilon_2  $\\
$( \bar \chi, N)$ & $+\phantom{\Big|}$
& $\bigl(\bone,\Ysymm\bigr)  $ & $ \sqrt{\chi_{ij}}   $\\
$(N_{\alpha a},D_{\alpha a})$ & $-\phantom{\Big|}$
& $\bigl(\bone,\Yasymm\bigr)  $ & $ \chi_{ij}+\epsilon_1+\epsilon_3,\, \chi_{ij}+\epsilon_1+\epsilon_4  $\\
$(w_{\dot \alpha},\mu_{\dot a})$ & $+\phantom{\Big|}$
& $\bigl(\Yfund,  {\Yfund}\bigr)  $
   & $ \chi_i-a_u+\ft{\epsilon_1+\epsilon_2}{2}    $  \\
   $(h_{a},\mu_{a})$ & $-\phantom{\Big|}$
& $\bigl( \Yfund,  {\Yfund} \bigr)  $
   &   $ \chi_i-a_u+\ft{\epsilon_3-\epsilon_4 }{2}  $  \\
\hline
\end{tabular}
\caption{Instanton moduli for the SO$(2N)$ gauge theory. The columns display the moduli, their statistics,
their transformation properties with respect to the gauge and instanton symmetry groups and their
$Q^2$-eigenvalues $\lambda_\phi$.  }
\label{tabson}
\end{center}
\end{table}

Collecting the eigenvalues $\lambda_\phi$ for all moduli, we find that the $k$-instanton partition function
is
\begin{equation}
Z_k=
\oint\prod_{i=1}^k\frac{d\chi_i}{2\pi\ii}~z_k^{\mathrm{gauge}}\,z_k^{\mathrm{matter}}
\label{Zkson}
\end{equation}
where
\begin{subequations}
\begin{align}
z_k^{\mathrm{gauge}}&= \frac{(-1)^k}{2^k\,k!}\,   \left( \frac{\epsilon_1+\epsilon_2    }{\epsilon_1\epsilon_2  }\right)^k
\,\frac{\Delta(0)\,\Delta(\epsilon_1+\epsilon_2 )}{\Delta(\epsilon_1)\,\Delta(\epsilon_2)}\,\prod_{i=1}^k
\frac{ 4\chi_i^2   \big( 4\chi_i^2-(\epsilon_1+\epsilon_2)^2\big)}{P\big(\chi_i+\frac{\epsilon_1+\epsilon_2 }{2}\big)P\big(\chi_i-\frac{\epsilon_1+\epsilon_2 }{2}\big)}
~,\\
&\notag\\
z_k^{\mathrm{matter}}&=\,\left( \frac{ (\epsilon_1+\epsilon_3)    (\epsilon_1+\epsilon_4)   }{   \epsilon_3 \epsilon_4 }\right)^{\!k}
\frac{\Delta\big(\epsilon_1+\epsilon_3 \big)\Delta\big(\epsilon_1+\epsilon_4 \big)  }  {\Delta\big(\epsilon_3 \big)\Delta\big(\epsilon_4 \big)  }
\prod_{i=1}^k
\frac{P\big(\chi_i+\frac{\epsilon_3-\epsilon_4}{2} \big)P\big(\chi_i- \frac{\epsilon_3-\epsilon_4}{2} \big)}{\big( 4 \chi_i^2-\epsilon_3^2 \big)\big( 4 \chi_i^2-\epsilon_4^2 \big)}
\end{align}
\end{subequations}
with
\begin{equation}
\begin{aligned}
P(x) =  \prod_{u=1}^N\big(x^2-a_u^2)~,\qquad
\Delta(x)=\prod_{i<j}^k\big((\chi_i-\chi_j)^2-x^2)\big)\big((\chi_i+\chi_j)^2-x^2\big)~.
\end{aligned}
\end{equation}
Once again the integrals in (\ref{Zkson}) are computed by closing the contours in the upper-half complex $\chi_i$-planes with the  prescription (\ref{prescription}).  It is important to stress that unlike in the U$(N)$ theory, the integral (\ref{Zkson}) receives non-trivial contributions also from poles located at $\chi_i=\epsilon_3$, $\chi_i=\epsilon_4$, $\chi_{ij}=\epsilon_3$ and $\chi_{ij}=\epsilon_4$.
The contributions of the corresponding residues are crucial to find an expression which is polynomial in the
hypermultiplet mass as one expects on general grounds.
Only at the very end of the computation one should use the identification (\ref{mass34}) in order to
write the final results in terms of the vacuum expectation values $a_u$ and the mass $m$ in the normalization
used in the previous sections.

\subsubsection*{1-instanton terms}
For $k=1$ the poles of (\ref{Zkson}) are located at
\begin{equation}
 \chi_1= \left\{   \pm a_u+\frac{\epsilon_1+\epsilon_2}{2}  ~,~\frac{\epsilon_3}{2}  ~,~\frac{\epsilon_4}{2} \right\}   \quad\mbox{for}~ u=1,\cdots, N
 \end{equation}
The $k=1$ prepotential can then be written as
\begin{equation}
F_{k=1}=-\epsilon_1 \epsilon_2 \, Z_1
=\sum_{u=1}^N f_{+a_u+\ft{\epsilon_1+\epsilon_2}{2}}+
\sum_{u=1}^N f_{-a_u+\ft{\epsilon_1+\epsilon_2}{2}}+
f_{\ft{\epsilon_3}{2}}+f_{\ft{\epsilon_4}{2}}
\label{Fk1SO2N}
\end{equation}
with
\begin{equation}
\begin{aligned}
f_{ \pm a_u+\ft{\epsilon_1+\epsilon_2}{2} } &=- (\epsilon_1+\epsilon_3) (\epsilon_1+\epsilon_4) \,
      \frac{(\pm 2a_u+\epsilon_1+\epsilon_2)(\pm a_u+\epsilon_1+\epsilon_2)}{
   (\pm 2a_u+\epsilon_1+\epsilon_2-\epsilon_3)( \pm 2a_u+\epsilon_1+\epsilon_2-\epsilon_4)}   \nonumber\\
&~~~~~~~~~\times    \prod_{v\neq u}  \frac{ \big( (\pm a_u-\epsilon_3)^2-a_v^2\big)
\big( (\pm a_u-\epsilon_4)^2-a_v^2\big)}{(a_u^2-a_v^2)
\big((\pm a_u+\epsilon_1+\epsilon_2)^2-a_v^2\big)  }~,\\
  f_{ \ft{\epsilon_3}{2} } &= -\frac{(\epsilon_1+\epsilon_3) (\epsilon_1+\epsilon_4) (\epsilon_3-\epsilon_1-\epsilon_2)} { 8 (\epsilon_3-\epsilon_4) }
      \prod_{u=1}^N  \frac{ (2\epsilon_3-\epsilon_4)^2-a_u^2}{(\epsilon_3-\epsilon)^2-a_u^2} ~,\\
f_{ \ft{\epsilon_4}{2} } &= -\frac{(\epsilon_1+\epsilon_3) (\epsilon_1+\epsilon_4) (\epsilon_4-\epsilon_1-\epsilon_2)}{8  (\epsilon_4-\epsilon_3) }
      \prod_{u=1}^N  \frac{(2\epsilon_4-\epsilon_3)^2-a_u^2}{(\epsilon_4-\epsilon)^2-a_u^2}  ~.
  \end{aligned}
\end{equation}
For example for SO(4) these formulas lead to
\begin{eqnarray}
F_{k=1}\Big|_{\mathrm{SO}(4)}&=&(M^2+h^2)\Big[-\frac{17}{8}+
\frac{M^2}{(a_1+a_2)(a_1+a_2+\epsilon)}+
\frac{M^2}{(-a_1-a_2)(-a_1-a_2+\epsilon)}\nonumber\\
&&\qquad+
\frac{M^2}{(a_1-a_2)(a_1-a_2+\epsilon)}+
\frac{M^2}{(-a_1+a_2)(-a_1+a_2+\epsilon)}
\Big]
\label{F1so4}
\end{eqnarray}
where we have used (\ref{mass34}), (\ref{M2}) and (\ref{epsilonh}).
Inside the square brackets the terms proportional to $M^2$ precisely reconstruct the sum $g_0$ defined in
(\ref{g2n}) so that this result is in perfect agreement with (\ref{FFk1a}).
We also notice that (\ref{F1so4}) is related to the SU(2) prepotential (\ref{Fk1SU2}). Indeed, upon comparison, we have
\begin{equation}
F_{k=1}\Big|_{\mathrm{SO}(4)}(a_1,a_2)= F_{k=1}\Big|_{\mathrm{SU}(2)}(a_L)+
F_{k=1}\Big|_{\mathrm{SU}(2)}(a_R)+
\frac{15}{8}(M^2+ h^2)
\end{equation}
where
\begin{equation}
a_1 =a_L+a_R~,    \qquad ~~~~~   a_2 =a_L-a_R ~,
\end{equation}
so the two prepotentials match up to an $a$-independent function as they should, since $\mathrm{SO}(4)\sim
\mathrm{SU}(2)\times\mathrm{SU}(2)$.

The explicit expressions of the prepotential for groups of higher rank quickly become rather involved; nevertheless we have checked up to SO(12) that the 1-instanton result (\ref{Fk1SO2N}) can be written as
\begin{equation}
\begin{aligned}
F_{k=1}&=M^2(M^2+h^2)\Big[g_0+\!\frac{M^2}{2}g_2+\!\frac{M^2(M^2-\epsilon^2)}{24}g_4
+\!\frac{M^2(M^4-3M^2\epsilon^2+3\epsilon^4)}{720}g_6+\!\cdots\!\Big]
\end{aligned}
\label{Fk1gn1}
\end{equation}
This is the same expression we found for the U$(N)$ theories (see (\ref{Fk1gn})) and is in perfect
agreement with what follows from solving the recursion relation as discussed in Section~\ref{sec:recepsi}.
Furthermore, in the limit $\epsilon,h\to0$ we exactly recover the results obtained in \cite{Ennes:1999fb} using
a very different approach.

\subsubsection*{2-instanton terms}
At $k=2$ one has to compute two integrals. Again, to avoid long formulas we only write an example
in the $\epsilon,h\to0$ limit for the purpose of illustration.
For SO(4), up to $a$-independent terms we find
\begin{equation}
\begin{aligned}
F_{k=2}\Big|_{\mathrm{SO}(4)}&=12m^4\,\frac{a_1^2+a_2^2}{\big(a_1^2-a_2^2\big)^2}-24m^6\,\frac{a_1^4+6a_1^2a_2^2+a_2^4}{\big(a_1^2-a_2^2\big)^4}\\
&~~~+
10m^8\frac{a_1^6+15a_1^4a_2^2+15a_1^2a_2^4+a_2^6}{\big(a_1^2-a_2^2\big)^6}\\
&=3m^4\, C_{2}-6m^6\,C_{4}+3m^6\,C_{2;11}+\frac{5m^8}{2}\,C_6+6m^8\,C_{4;2}+\frac{m^8}{2}\,C_{3;3}+\frac{m^8}{2}\,C_{2;1111}
\end{aligned}
\end{equation}
where the last line follows upon using the sums (\ref{defcnms}) over the lattice root of SO$(2N)$.
Formally, this is the same expression found for the unitary theories and agrees with the results obtained in Section~\ref{secn:recursion} from the recursion relations.

We have verified that this agreement persist up for higher rank groups up to SO(12). This fact puts
our findings on a very solid ground.

\section{Conclusions}
\label{sec:concl}
In this paper we have shown that S-duality in $\mathcal{N}=2^\star$ gauge theories with simply-laced gauge groups requires that the quantum prepotential
satisfies a modular anomaly equation which in turn allows to recursively determine the prepotential itself.
It is very satisfactory that these conditions can be expressed in a unified form involving sums over the roots of the gauge algebra without resorting to the specific details of the algebra itself. This is the key to extend our results to the case of exceptional algebras, where the lack of an ADHM construction of the instanton moduli space does not allow the application of the traditional methods of investigation. The differential equation coming from the anomaly, irrespective of the gauge algebra, needs an external output to fix all the terms in the prepotential. Given that $f_n$ is a modular form of weight $2n-2$, in solving the recursion relation (\ref{rec}) we can add to $f_n$ all monomials in the Eisenstein series which have weight $2n-2$ but do not
contain $E_2$. The coefficients in front of these terms are determined by comparing
with the perturbative expansion and when this is not enough by resorting to microscopic instanton computations. Given that no microscopic instanton computations exist for the exceptional
gauge groups this could seem a problem.
Luckily enough,  the results for $f_n$ given in terms of sums over the roots of the algebra
are universal and thus should hold for the exceptional algebras as well.
We believe our results be a very solid conjecture which we have successfully tested for the lowest instanton number with the result for pure $\mathcal{N}=2$ theory existing in literature \cite{Keller:2011ek,Keller:2012da} and provide an elegant generalisation to the $\mathcal{N}=2^\star$  case, as well as precise predictions for higher instanton corrections.

\vskip 1.5cm
\noindent {\large {\bf Acknowledgments}}
\vskip 0.2cm
We thank Carlo Angelantonj, Sujay Ashok, Eleonora Dell'Aquila and Igor Pesando for discussions.

The work of M.B., M.F. and A.L. is partially supported  by the Compagnia di San Paolo
contract ``MAST: Modern Applications of String Theory'' TO-Call3-2012-0088.

\vskip 1cm

\appendix
\section{Notations and conventions for the root systems}
\label{secn:roots}
In this appendix we list our conventions for the root systems of the simply-laced algebras. 
We consider both the classical algebras $A_r=\mathrm{su}(r+1)$ and $D_r= \mathrm{so}(2r)$, and 
the exceptional ones $E_6$, $E_7$ and $E_8$.

We denote by $\Psi$ the set of all roots $\alpha$ and
by $\Psi(\alpha)$ the set
\begin{equation}
\begin{aligned}
\Psi(\alpha)&=\left\{\beta\in\Psi \, : \,\alpha\cdot\beta=1\right\}~.
\end{aligned}
\end{equation}
The order of this set is 
\begin{equation}
\mathrm{ord}\big( \Psi(\alpha)\big)= 2h^{\!\vee}-4~,\\
\label{dimCn}
\end{equation}
where $h^{\!\vee}$ is the dual Coxeter number of the algebra. 

To write the roots of the different ADE algebras we use 
the standard orthonormal basis in $\mathbb{R}^r$: $\{\mathbf{e}_i\,;\,1\le i\le r\}$
and our conventions are such that, for every ADE algebra 
$(\alpha \cdot \alpha)=2$.

\subsection*{The roots of $A_r$}
\label{subsecn:bn}
The set $\Psi$ of the roots of $A_r$ is 
\begin{equation}
\Psi=\big\{\pm (\mathbf{e}_i - \mathbf{e}_j) \,;\,1\le i<j\le r\big\}~.
\label{rAn}
\end{equation}
It is easy to see that 
\begin{equation}
\mathrm{ord}\big(\Psi \big)= r(r-1) \quad{\rm and}\quad
\mathrm{ord}\big( \Psi(\alpha)\big)= 2r-4 ~,
\label{dimensAn}
\end{equation}
since the dual Coxeter number for $A_r$ is $h^{\!\vee}=r$.

\subsection*{The roots of $D_r$}
For $D_r$ the roots are given by
\begin{equation}
\Psi=\big\{\pm \mathbf{e}_i\pm \mathbf{e}_j\,;\,1\le i<j\le r\big\}~,
\label{rDn}
\end{equation}
with all possible signs. It is easy to see that 
\begin{equation}
\mathrm{ord}\big(\Psi\big)= 2r(r-1) \quad{\rm and}\quad
\mathrm{ord}\big( \Psi(\alpha)\big)=4r-8~,
\label{dimensDn}
\end{equation}
since the dual Coxeter number for $D_r$ is $h^{\!\vee}=2r-2$.

\subsection*{The roots of $E_6$}
$E_6$ has $72$ roots given by
\begin{equation}
\Psi=\big\{\pm \mathbf{e}_i\pm \mathbf{e}_j\,;\,1\le i<j\le 5\big\}
\cup
\big\{\ft{1}{2}\big(\pm \mathbf{e}_1\pm
\cdots \pm \mathbf{e}_5
\pm \sqrt{3}\, \mathbf{e}_6\big)
\, \big\}~,
\label{rE6}
\end{equation}
where the elements of the second set must have an even number of minus signs. 
In this case
\begin{equation}
\mathrm{ord}\big( \Psi(\alpha)\big)=20~,\quad
\label{dimE6}
\end{equation}
since $h^{\!\vee}=12$.

\subsection*{The roots of $E_7$}
The 126 roots of $E_7$ are
\begin{equation}
\Psi=\big\{\pm \mathbf{e}_i\pm \mathbf{e}_j\,;\,1\le i<j\le 6\big\}
\cup \big\{\pm\sqrt{2}\,\mathbf{e}_7\big\}
\cup
\big\{\ft{1}{2}\big(\pm \mathbf{e}_1\pm 
\cdots  \mathbf{e}_6 \pm \sqrt{2} \mathbf{e}_7\big)\, \big\}~,
\label{rE7}
\end{equation}
where the elements of the third set must have an odd (even) number of minus signs in the 
$(\mathbf{e}_1, \cdots, \mathbf{e}_6)$ components if the $\mathbf{e}_7$ is positive (negative).
Moreover we have
\begin{equation}
\mathrm{ord}\big( \Psi(\alpha)\big)=32~,\quad
\label{dimE7}
\end{equation}
since in this case $h^{\!\vee}=18$.

\subsection*{The roots of $E_8$}
$E_8$ has $240$ roots given by
\begin{equation}
\Psi=\big\{\pm \mathbf{e}_i\pm \mathbf{e}_j\,;\,1\le i<j\le 8\big\}
\cup
\big\{\ft{1}{2}\big(
\pm \mathbf{e}_1\pm \cdots  \mathbf{e}_7
\pm \mathbf{e}_8\big)
\, \big\}~,
\label{rE8}
\end{equation}
where the element of the second set must have an even number of minus signs.
It is easy to see that that
\begin{equation}
\mathrm{ord}\big( \Psi(\alpha)\big)=56~,\quad
\label{dimE8}
\end{equation}
since in this case $h^{\!\vee}=30$.
\section{Eisenstein series and their modular properties}
\label{appb}
The Eisenstein series $E_{2n}$ are holomorphic functions of $\tau$ defined as
\begin{equation}
\label{defeis}
E_{2n} = \frac{1}{2\zeta(2n)}\sum_{m,n\in\mathbb{Z}^2\setminus \{0,0\}} \frac{1}{(m+n\tau)^{2n}}~.
\end{equation}
For $n>1$, they are modular forms of degree $2n$: under an $\mathrm{SL}(2,\mathbb{Z})$ transformation
\begin{equation}
\label{sl2z}
\tau \to \tau^\prime = \frac{a\tau + b}{c\tau +d}~~~\mbox{with}~~
a,b,c,d \in\mathbb{Z}~~~\mbox{and}~~
ad-bc=1
\end{equation}
one has
\begin{equation}
\label{eissl2z}
E_{2n}(\tau^\prime) = (c\tau + d)^{2n} E_{2n}(\tau)~.
\end{equation}
For $n=1$, the $E_2$ series is instead quasi-modular:
\begin{equation}
\label{eis2sl2z}
E_{2}(\tau^\prime) = (c\tau + d)^{2} E_{2}(\tau) + \frac{6}{\ii\pi}c(c\tau + d)~.
\end{equation}

All the modular forms of degree $2n>6$ can be expressed in terms of $E_4$ and $E_6$; the
quasi-modular forms instead can be expressed as polynomials in $E_2$, $E_4$ and $E_6$.

The Eisenstein series admit a Fourier expansion in terms of $q=\ee^{2\pi\ii\tau)}$ of the form 
\begin{equation}
E_{2n}=1+\frac{2}{\zeta(1-2n)}\sum_{k=1}^\infty \sigma_{2n-1}(k) q^{k}~,
\label{eis2nq0}
\end{equation}
where $\sigma_p(k)$ is the sum of the $p$-th powers of the divisors of $k$. 
In particular, this amounts to
\begin{equation}
\label{e246q}
\begin{aligned}
E_2 & = 1 - 24 \sum_{k=1}^\infty \sigma_1(k) q^{k} = 1 - 24 q - 72 q^2 - 96 q^3 + \ldots~,\\
E_4 & = 1 + 240 \sum_{k=1}^\infty \sigma_3(k) q^{k} = 1 + 240 q + 2160 q^2 + 6720 q^3 + \ldots~,\\
E_6 & = 1 - 504 \sum_{k=1}^\infty \sigma_5(k) q^{k} =1 - 504 q - 16632 q^2 - 122976 q^3 + \ldots~.
\end{aligned}
\end{equation}
Using these expansions it is easy to see that a generic quasi-modular function of weight $4$ 
\begin{equation}
M_4(q) = \alpha E_2^2 + \beta E_4 = (\alpha+\beta) + 48 (5\beta - \alpha) q + \cdots
\label{gg44}
\end{equation}
has no perturbative contribution if $\alpha=-\beta$, like for instance $E_2^2-E_4$, and it has no 
$1$-instanton contribution if $\alpha=5\beta$, like for instance $5E_2^2+E_4$. These are precisely the
combinations that appear in the prepotential coefficient $f_3$ (see (\ref{f3tris}).
Analogously, for a weight $6$ quasi-modular function
\begin{equation}
M_6(q) = \alpha E_2^3 + \beta E_2E_4 +\gamma E_6= (\alpha+\beta+\gamma) + 
72 (- \alpha+3 \beta -7\gamma) q + \cdots
\label{gg66}
\end{equation}
the perturbative and $1$-instanton contributions vanishes respectively for $\alpha=-\beta-\gamma$
and $\alpha=3\beta-7\gamma$. This is the case for the combinations that appear in the
prepotential coefficient $f_4$ in (\ref{f5is}).

The $E_2$ series is related to the Dedekind $\eta$-function
\begin{equation}
\label{etadef}
\eta(q) = q^{1/24}\prod_{k=1}^\infty (1 - q^{k}).
\end{equation}
Indeed, taking the logarithm of this definition we get
\begin{equation}
\label{logeta}
\log\left(\frac{\eta(q)}{q^{1/24}}\right) = \sum_{r=1}^\infty \log\left(1 - q^{r}\right)
= - \sum_{k=1}^\infty \frac{\sigma_1(k)}{k} q^{k}~.
\end{equation}
If we apply now to this relation the derivative operator $q \frac{d}{dq}$ we get
\begin{equation}
\label{Dlogeta}
q \frac{d}{dq} \log\left(\frac{\eta}{q^{1/24}}\right) = - \sum_{k=1}^\infty \sigma_1(k) q^{k} =
\frac{E_2 - 1}{24}~. 
\end{equation}
Applying repeatedly the operator $q \frac{d}{dq}$ to this last expression we also find
\begin{equation}
\label{repDe2}
(q \frac{d}{dq})^{n-1} (E_2 -1) = -24 \sum_{k=1}^\infty k^{n-1} \sigma_1(k) q^{k}~.
\end{equation}
Finally, one has
\begin{equation}
\label{De2e4d6}
\begin{aligned}
q \frac{d}{dq} E_2 & = \frac{1}{12} \left(E_2^2 - E_4\right)~,\\
q \frac{d}{dq} E_4 & = \frac{1}{3}  \left(E_2 E_4 - E_6\right)~,\\
q \frac{d}{dq} E_6 & = \frac{1}{2}  \left(E_2 E_6 - E_4^2\right)~.
\end{aligned}
\end{equation}

\section{The $\Gamma_2$-function}
\label{secn:appgamma2}
The Barnes double $\Gamma$-function is defined as
\begin{eqnarray}
\label{deflg2}
&&\log\Gamma_2(x|\epsilon_1,\epsilon_2)  = \frac{d}{ds}\left(  \frac{\Lambda^s}{\Gamma(s) } \int_0^\infty \frac{dt}{t}
\frac{t^s\, \ee^{-x t}}{(1-\ee^{-\epsilon_1 t} ) (1-\ee^{-\epsilon_2 t} )}\right)\Big|_{s=0}
\label{ge1e2}\\
&&~~~~~= \log \left( \frac{x}{\Lambda}\right)^2\Big(-\frac{1}{4} b_0 \, x^2  +\frac{1}{2}  b_1 \, x -\frac{b_2}{4}\Big)
+  \Big( \frac{3}{4} b_0 \, x^2 -b_1 \, x    \Big)  +\sum_{n=3}^\infty  \frac{b_n x^{2-n}}{n(n-1)(n-2)}
\nonumber
\end{eqnarray}
where the coefficients  $b_n$ are given by 
\begin{equation}
\label{cndef}
\frac{1}{(1-\ee^{-\epsilon_1 t} ) (1-\ee^{-\epsilon_2 t} ) }=\sum_{n=0}^\infty  \frac{b_n}{n!}\, t^{n-2}~.
\end{equation}
The first few of them are
\begin{equation}
\label{cnfirst}
b_0=\frac{1}{\epsilon_1 \epsilon_2}=\frac{1}{h^2}~,  \qquad   
b_1=\frac{\epsilon_1+\epsilon_2}{2 \epsilon_1 \epsilon_2}=\frac{\epsilon}{h^2}~,\qquad b_2=\frac{\epsilon_1^2+3 \epsilon_1 \epsilon_2+\epsilon_2^2}{6\epsilon_1 \epsilon_2}=\frac{4\epsilon^2+h^2}{6h^2}~.
\end{equation}
\section{Useful formulas for the root lattice sums}
\label{secn:sums}
Let us first recall the definition of the lattice sums $C_{n;m_1m_2\cdots}$ introduced 
in (\ref{defcnms}) which we rewrite here for convenience:
\begin{equation}
\label{defci}
C_{n;m_1m_2\ldots m_\ell} = \sum_{\alpha\in\Psi}
\sum_{\beta_1\not=\beta_2\not=\ldots\beta_\ell\in\Psi(\alpha)} 
\frac{1}{(\alpha\cdot a)^n(\beta_1\cdot a)^{m_1}(\beta_2\cdot a)^{m_2}\cdots (\beta_\ell\cdot a)^{m_\ell}}
\end{equation}
where $\Psi$ is the root lattice and $\Psi(\alpha)$ is the set :
\begin{equation}
\label{defDeltaalphaapp}
\Psi(\alpha) = \left\{\beta\in\Psi:~ \alpha\cdot\beta=1 \right\}~.
\end{equation}
Actually, not all these sums are independent of each other, since there exist 
various algebraic identities among them which we are going to discuss.

First of all, from the definition (\ref{defci}) it is straightforward to see that
\begin{equation}
C_n=0\quad\mbox{for~$n$~odd}~,
\label{Cnodd}
\end{equation}
and, more generally
\begin{equation}
C_{n;m_1m_2\ldots,m_\ell}=0\quad\mbox{for~$\big(n+\sum_\ell m_\ell\big)$~odd}~.
\label{Cnmodd}
\end{equation}
Many different sets of relations among the $C$'s can be proved by dividing the sums in 
(\ref{defci}) into sums over closed orbits of the Weyl reflection group and exploiting the properties of the partial sums. As a first example of this strategy, let us consider the set formed by 
a given couple of roots $(\alpha,\beta)$ with
$\beta \in\Psi(\alpha)$, together with its images under mutual reflection:
\begin{equation}
\label{Worbit2}
\big\{\,(\alpha,\,\beta)~,~ (-\alpha,\,\beta-\alpha)~,~ (\alpha-\beta\,,-\beta)\,\big\}~.
\end{equation} 
This set forms a closed orbit of the Weyl reflections group, up to irrelevant overall signs.
It is straightforward to see that
\begin{equation}
\label{sumor2}
\frac{1}{(\alpha\cdot a)(\beta\cdot a)} + \frac{1}{(\alpha\cdot a)
((\alpha-\beta)\cdot a)}
- \frac{1}{((\alpha-\beta)\cdot a)(\beta\cdot a)} = 0~.
\end{equation}
{From} this equation we can immediately prove that
\begin{equation}
C_{1;1}=
\sum_{\alpha\in\Psi} \sum_{\beta\in\Psi(\alpha)} \frac{1}{(\alpha\cdot a)(\beta\cdot a)} =
0~.
\label{C11a}
\end{equation}
In fact, the sum in $C_{1;1}$ contains (twice) all the images of any couple under Weyl reflections.
The expression in the above sum vanishes identically already when summed over the components
of the set in (\ref{Worbit2}).

The implication of (\ref{sumor2}) are however far more reaching. In fact, multiplying 
it by $1/(\alpha\cdot a)^2$ and then summing over $\alpha\in \Psi$ and $\beta \in \Psi(\alpha) $, 
we easily find
\begin{equation}
2\,C_{3;1}=\widehat C_{2;11}~,
\label{C31}
\end{equation}
with
\begin{equation}
\widehat C_{2;11}= \sum_{\alpha\in\Psi}\sum_{\beta\in\Psi(\alpha)} 
            \frac{1}{(\alpha\cdot a)^2(\beta\cdot a)((\alpha-\beta)\cdot a)}~.
\end{equation}
Likewise, multiplying (\ref{sumor2}) by $1/(\alpha\cdot a)(\beta\cdot a)$ 
and then summing over the roots, we find
\begin{equation}
C_{2;2}= - 2\,\widehat C_{2;11}~.
\label{C22}
\end{equation}
These two relations together imply
\begin{equation}
C_{2;2}=-4\,C_{3;1}~,
\label{C22C31}
\end{equation}
but, using again the strategy of summing over the closed orbits of the Weyl group, 
it is easy to check that
\begin{equation}
\widehat C_{2;11}= C_{2;11}~,
\end{equation}
thus obtaining the identity 
\begin{equation}
2\,C_{3;1}= C_{2;11}~.
\label{C31C211}
\end{equation}
This is the identity used to write the expression (\ref{f3is}) for $f_3$ given in the main text.

This method can be easily generalized to derive many other identities.
In fact, multiplying (\ref{sumor2}) by 
\begin{equation}
 \frac{1}{(\alpha\cdot a)^{n-1}(\beta\cdot a)^{m-1}}
\end{equation}
and summing over the roots, one gets
\begin{equation}
C_{n;m}= -\widehat C_{n;m-1 \,p}-\widehat C_{n-1;m \,p}~.
\label{C1}
\end{equation}
This relation, together with the symmetry properties of the different $C$'s, can be recursively
used to link together all lattice sums with two indices at a given level. For example, at level 6
we have
\begin{equation}
4C_{5;1} + 2C_{4;2}+ C_{3;3}=0 ~.
\label{Csix}
\end{equation}
All identities among the root lattice sums, presumably, can be derived following the
strategy we have outlined; however, as the level increases, the algebra becomes 
more involved and thus it is often more convenient to check the identities in a numerical way.

Here we list some other identities among the level 6 sums that are needed 
in order to write the expression for $f_4$ as in \eq{f4fin}, namely
\begin{equation}
\begin{aligned}
C_{4;11} &= -C_{4;2}-\frac{1}{2}\,C_{3;3} +\frac{1}{12} \,C_{2;1111} ~,\qquad
C_{3;21} = -\frac{1}{2} \,C_{3;3} - \frac{1}{4}\,C_{2;1111}~,\\
C_{2;31} &= \frac{1}{2}\, C_{3;3} ~,\qquad C_{3;111} =\frac{1}{2}\, C_{2;1111}~,\qquad
C_{2;211} =-\frac{2}{3}\, C_{2;1111}~.
\end{aligned}
\label{identfin}
\end{equation}
\section{An observation on the prepotential}
\label{app:observation}
It is a fact that in all prepotential coefficients $f_n$ the combinations of Eisenstein series 
appearing in front of the sum $C_{2;11\cdots1}/\ell!$ have a linear term in $q$ whose coefficient
is exactly 1. This comes about through the following mechanism. 
{From} (\ref{f2fin}) we see that the coefficient of $C_2$ in $f_2$ is (for simplicity we omit now the overall powers of $m$ which 
are easily reinstated)
\begin{equation}
\label{f2C2q}
- \frac{E_2}{24} = -\frac{1}{24} + q + \cdots~,
\end{equation}   
so it enjoys this property. 
The coefficient of $C_{2;11}/2$ in $f_3$ (see (\ref{f3fin})), of $C_{2;1111}/24$ in $f_4$ (see (\ref{f4fin})), and so on,  turn out to be obtained from (\ref{f2C2q}) by repeated 
logarithmic $q$-derivatives; indeed using (\ref{De2e4d6}) one has
\begin{equation}
\label{f3C211q}
- \frac{E_2}{24} ~\stackrel{~~q\frac{d\,}{dq}}{\phantom{\big|}\longrightarrow}~ -\frac{E_2^2 - E_4}{288}
~\stackrel{~~q\frac{d\,}{dq}}{\phantom{\big|}\longrightarrow}~ -\frac{E_2^3 - 3 E_2 E_4 + 2 E_6}{1728} 
~\stackrel{~~q\frac{d\,}{dq}}{\phantom{\big|}\longrightarrow}~
\cdots
\end{equation} 
so they again contain exactly $q$ in their Fourier expansion. 
Due to this structure, the entire part of the prepotential which contains the expressions 
$C_{2;11\ldots}/\ell!$, and not just its 1-instanton component, can be written in a compact way, analogously 
to (\ref{f1qnop}). Reinstating the mass prefactors and introducing the operator
\begin{equation}
\label{defD}
\cD = m^2\, q\frac{d~}{dq} = \frac{m^2}{2\pi\ii}\,\frac{d\,}{d\tau}
\end{equation}
such terms are  
\begin{equation}
\label{c2211series}
\begin{aligned}
& \Big(C_2 + \frac{1}{2!}\, C_{2;11} \cD + \frac{1}{4!}\, C_{2;1111} \cD^2 +
\frac{1}{6!}\,C_{2;111111}\cD^3 + \cdots\Big) \Big(\!-\frac{E_2}{24}\,\Big)\\
& \qquad= \sum_{\alpha\in\Psi} \frac{1}{(\alpha\cdot a)^2} 
\prod_{\beta\in\Psi(\alpha)}
\Big(1 + \frac{\cD^{\frac{1}{2}}}{\beta\cdot a}\Big)\Big(\!-\frac{E_2}{24}\,\Big)~.
\end{aligned}
\end{equation}
Notice that only integer powers of $\cD$ remain in the expansion of the product above, due to the properties
of the root systems.

Actually, all terms in the prepotential can be grouped into series of terms connected by the action of the operator $\cD$; 
writing only the terms up to $f_4$ for simplicity, we may write
\begin{equation}
\label{resf2f3}
\begin{aligned}
f_2 & = \cA_{(1)}\,C_2~,\\
f_3 & = \cA_{(2)}\, C_4 + \cD \cA_{(1)} \,\frac{1}{2} C_{2,1,1}~,\\ 
f_4 & = \cA_{(3)}\, C_6 - \cD \cA_{(2)} \, \frac{1}{2}  \left(C_{4,2} +\frac{1}{12} C_{3,3}\right)
+  \cD^2 \cA_{(1)}\,\frac{1}{24}C_{2,1,1,1,1}~,
\end{aligned}
\end{equation}
where
\begin{equation}
\label{calA12}
\begin{aligned}
\cA_{(1)} &= -\frac{m^4}{24} E_2~,\\
\cA_{(2)} &= -\frac{m^6}{720} \left(5E_2^2 + E_4\right)~,\\
\cA_{(3)} &= - \frac{m^8}{90720} \left(175 E_2^3 + 84 E_2 E_4 + 11 E_6\right)~.
\end{aligned}
\end{equation}
Thus, at each order $m^{2n}$, a new modular form $\cA_{(n)}$ appears in front of the structure
$C_n$. For $n>2$, the term of order $q$ in this form vanishes; this agrees with the fact that in the microscopic
computation of the 1-instanton corrections the structure $C_n$ cannot appear. 
All the rest of $f_n$ is organized in structures proportional to multiple $\cD$ derivatives of the 
lower $\cA_{(k)}$ coefficients, which have therefore neither perturbative terms (since the $q$-derivative kills the constant term in $\cA_{(k)}$) nor 1-instanton corrections, 
except for the structures $C_{2;11\cdots}$, proportional to derivatives 
of $\cA_{(1)}$, that we considered above.

\providecommand{\href}[2]{#2}\begingroup\raggedright\endgroup


\begin{thebibliography}{10}

\bibitem{Montonen:1977sn}
C.~Montonen and D.~I. Olive, \emph{{Magnetic monopoles as gauge particles?}},
\href{http://dx.doi.org/10.1016/0370-2693(77)90076-4}{Phys. Lett. {\bf B72}
  (1977)  117}.

\bibitem{Vafa:1994tf}
C.~Vafa and E.~Witten, \emph{{A Strong coupling test of S duality}},
  \href{http://dx.doi.org/10.1016/0550-3213(94)90097-3}{Nucl. Phys. {\bf B431}
  (1994)  3--77},
\href{http://arxiv.org/abs/hep-th/9408074}{{\tt arXiv:hep-th/9408074
  [hep-th]}}.

\bibitem{Girardello:1995gf}
L.~Girardello, A.~Giveon, M.~Porrati, and A.~Zaffaroni, \emph{{S duality in N=4
  Yang-Mills theories with general gauge groups}},
  \href{http://dx.doi.org/10.1016/0550-3213(95)00177-T}{Nucl. Phys. {\bf B448}
  (1995)  127--165},
\href{http://arxiv.org/abs/hep-th/9502057}{{\tt arXiv:hep-th/9502057
  [hep-th]}}.

\bibitem{Goddard:1976qe}
P.~Goddard, J.~Nuyts, and D.~I. Olive, \emph{{Gauge theories and magnetic
  charge}},
\href{http://dx.doi.org/10.1016/0550-3213(77)90221-8}{Nucl. Phys. {\bf B125}
  (1977)  1}.

\bibitem{Seiberg:1994rs}
N.~Seiberg and E.~Witten, \emph{{Electric - magnetic duality, monopole
  condensation, and confinement in N=2 supersymmetric Yang-Mills theory}},
  \href{http://dx.doi.org/10.1016/0550-3213(94)90124-4,
  10.1016/0550-3213(94)90124-4}{Nucl.Phys. {\bf B426} (1994)  19--52},
\href{http://arxiv.org/abs/hep-th/9407087}{{\tt arXiv:hep-th/9407087
  [hep-th]}}.

\bibitem{Seiberg:1994aj}
N.~Seiberg and E.~Witten, \emph{{Monopoles, duality and chiral symmetry
  breaking in N=2 supersymmetric QCD}},
  \href{http://dx.doi.org/10.1016/0550-3213(94)90214-3}{Nucl. Phys. {\bf B431}
  (1994)  484--550},
\href{http://arxiv.org/abs/hep-th/9408099}{{\tt arXiv:hep-th/9408099}}.

\bibitem{Nekrasov:2002qd}
N.~Nekrasov, \emph{{Seiberg-Witten prepotential from instanton counting}}, Adv.
  Theor. Math. Phys. {\bf 7} (2004)  831--864,
\href{http://arxiv.org/abs/hep-th/0206161}{{\tt arXiv:hep-th/0206161}}.

\bibitem{Flume:2002az}
R.~Flume and R.~Poghossian, \emph{{An algorithm for the microscopic evaluation
  of the coefficients of the Seiberg-Witten prepotential}},
  \href{http://dx.doi.org/10.1142/S0217751X03013685}{Int. J. Mod. Phys. {\bf
  A18} (2003)  2541},
\href{http://arxiv.org/abs/hep-th/0208176}{{\tt arXiv:hep-th/0208176}}.

\bibitem{Nekrasov:2003rj}
N.~Nekrasov and A.~Okounkov, \emph{{Seiberg-Witten theory and random
  partitions}},
\href{http://arxiv.org/abs/hep-th/0306238}{{\tt arXiv:hep-th/0306238}}.

\bibitem{Bruzzo:2002xf}
U.~Bruzzo, F.~Fucito, J.~F. Morales, and A.~Tanzini, \emph{{Multi-instanton
  calculus and equivariant cohomology}}, JHEP {\bf 05} (2003)  054,
\href{http://arxiv.org/abs/hep-th/0211108}{{\tt arXiv:hep-th/0211108}}.

\bibitem{Fucito:2004ry}
F.~Fucito, J.~F. Morales, and R.~Poghossian, \emph{{Multi instanton calculus on
  ALE spaces}},
  \href{http://dx.doi.org/10.1016/j.nuclphysb.2004.09.014}{Nucl.Phys. {\bf
  B703} (2004)  518--536},
\href{http://arxiv.org/abs/hep-th/0406243}{{\tt arXiv:hep-th/0406243
  [hep-th]}}.

\bibitem{Gaiotto:2012uq}
D.~Gaiotto and S.~S. Razamat, \emph{{Exceptional indices}},
  \href{http://dx.doi.org/10.1007/JHEP05(2012)145}{JHEP {\bf 05} (2012)  145},
\href{http://arxiv.org/abs/1203.5517}{{\tt arXiv:1203.5517 [hep-th]}}.

\bibitem{Keller:2012da}
C.~A. Keller and J.~Song, \emph{{Counting Exceptional Instantons}},
  \href{http://dx.doi.org/10.1007/JHEP07(2012)085}{JHEP {\bf 1207} (2012)
  085},
\href{http://arxiv.org/abs/1205.4722}{{\tt arXiv:1205.4722 [hep-th]}}.

\bibitem{Benvenuti:2010pq}
S.~Benvenuti, A.~Hanany, and N.~Mekareeya, \emph{{The Hilbert series of the one
  instanton moduli space}},
  \href{http://dx.doi.org/10.1007/JHEP06(2010)100}{JHEP {\bf 06} (2010)  100},
\href{http://arxiv.org/abs/1005.3026}{{\tt arXiv:1005.3026 [hep-th]}}.

\bibitem{Keller:2011ek}
C.~A. Keller, N.~Mekareeya, J.~Song, and Y.~Tachikawa, \emph{{The ABCDEFG of
  Instantons and W-algebras}},
  \href{http://dx.doi.org/10.1007/JHEP03(2012)045}{JHEP {\bf 1203} (2012)
  045},
\href{http://arxiv.org/abs/1111.5624}{{\tt arXiv:1111.5624 [hep-th]}}.

\bibitem{Hanany:2012dm}
A.~Hanany, N.~Mekareeya, and S.~S. Razamat, \emph{{Hilbert Series for Moduli
  Spaces of Two Instantons}},
  \href{http://dx.doi.org/10.1007/JHEP01(2013)070}{JHEP {\bf 1301} (2013)
  070},
\href{http://arxiv.org/abs/1205.4741}{{\tt arXiv:1205.4741 [hep-th]}}.

\bibitem{Cremonesi:2014xha}
S.~Cremonesi, G.~Ferlito, A.~Hanany, and N.~Mekareeya, \emph{{Coulomb branch
  and the moduli space of instantons}},
  \href{http://dx.doi.org/10.1007/JHEP12(2014)103}{JHEP {\bf 12} (2014)  103},
\href{http://arxiv.org/abs/1408.6835}{{\tt arXiv:1408.6835 [hep-th]}}.

\bibitem{Teschner:2014oja}
J.~Teschner, \emph{{Exact results on N=2 supersymmetric gauge theories}},
\href{http://arxiv.org/abs/1412.7145}{{\tt arXiv:1412.7145 [hep-th]}}.

\bibitem{Minahan:1997if}
J.~A. Minahan, D.~Nemeschansky, and N.~P. Warner, \emph{{Instanton expansions
  for mass deformed N=4 superYang-Mills theories}},
  \href{http://dx.doi.org/10.1016/S0550-3213(98)00314-9}{Nucl. Phys. {\bf B528}
  (1998)  109--132},
\href{http://arxiv.org/abs/hep-th/9710146}{{\tt arXiv:hep-th/9710146
  [hep-th]}}.

\bibitem{Bershadsky:1993ta}
M.~Bershadsky, S.~Cecotti, H.~Ooguri, and C.~Vafa, \emph{{Holomorphic anomalies
  in topological field theories}},
  \href{http://dx.doi.org/10.1016/0550-3213(93)90548-4}{Nucl.Phys. {\bf B405}
  (1993)  279--304},
\href{http://arxiv.org/abs/hep-th/9302103}{{\tt arXiv:hep-th/9302103
  [hep-th]}}.

\bibitem{Witten:1993ed}
E.~Witten, \emph{{Quantum background independence in string theory}},
\href{http://arxiv.org/abs/hep-th/9306122}{{\tt arXiv:hep-th/9306122
  [hep-th]}}.

\bibitem{Aganagic:2006wq}
M.~Aganagic, V.~Bouchard, and A.~Klemm, \emph{{Topological strings and (almost)
  modular forms}},
  \href{http://dx.doi.org/10.1007/s00220-007-0383-3}{Commun.Math.Phys. {\bf
  277} (2008)  771--819},
\href{http://arxiv.org/abs/hep-th/0607100}{{\tt arXiv:hep-th/0607100
  [hep-th]}}.

\bibitem{Gunaydin:2006bz}
M.~Gunaydin, A.~Neitzke, and B.~Pioline, \emph{{Topological wave functions and
  heat equations}}, \href{http://dx.doi.org/10.1088/1126-6708/2006/12/070}{JHEP
  {\bf 0612} (2006)  070},
\href{http://arxiv.org/abs/hep-th/0607200}{{\tt arXiv:hep-th/0607200
  [hep-th]}}.

\bibitem{Huang:2006si}
M.-x. Huang and A.~Klemm, \emph{{Holomorphic anomaly in gauge theories and
  matrix models}}, \href{http://dx.doi.org/10.1088/1126-6708/2007/09/054}{JHEP
  {\bf 0709} (2007)  054},
\href{http://arxiv.org/abs/hep-th/0605195}{{\tt arXiv:hep-th/0605195
  [hep-th]}}.

\bibitem{Grimm:2007tm}
T.~W. Grimm, A.~Klemm, M.~Marino, and M.~Weiss, \emph{{Direct integration of
  the topological string}},
  \href{http://dx.doi.org/10.1088/1126-6708/2007/08/058}{JHEP {\bf 0708} (2007)
   058},
\href{http://arxiv.org/abs/hep-th/0702187}{{\tt arXiv:hep-th/0702187
  [hep-th]}}.

\bibitem{Huang:2009md}
M.-x. Huang and A.~Klemm, \emph{{Holomorphicity and modularity in
  Seiberg-Witten theories with matter}},
  \href{http://dx.doi.org/10.1007/JHEP07(2010)083}{JHEP {\bf 1007} (2010)
  083},
\href{http://arxiv.org/abs/0902.1325}{{\tt arXiv:0902.1325 [hep-th]}}.

\bibitem{Huang:2010kf}
M.-x. Huang and A.~Klemm, \emph{{Direct integration for general $\Omega$
  backgrounds}},
\href{http://arxiv.org/abs/1009.1126}{{\tt arXiv:1009.1126 [hep-th]}}.

\bibitem{Huang:2011qx}
M.-x. Huang, A.-K. Kashani-Poor, and A.~Klemm, \emph{{The $\Omega$ deformed
  B-model for rigid $\mathcal{N}=2$ theories}},
  \href{http://dx.doi.org/10.1007/s00023-012-0192-x}{Annales Henri Poincare
  {\bf 14} (2013)  425--497},
\href{http://arxiv.org/abs/1109.5728}{{\tt arXiv:1109.5728 [hep-th]}}.

\bibitem{Galakhov:2012gw}
D.~Galakhov, A.~Mironov, and A.~Morozov, \emph{{S-duality as a beta-deformed
  Fourier transform}}, \href{http://dx.doi.org/10.1007/JHEP08(2012)067}{JHEP
  {\bf 1208} (2012)  067},
\href{http://arxiv.org/abs/1205.4998}{{\tt arXiv:1205.4998 [hep-th]}}.

\bibitem{Billo:2013fi}
M.~Billo, M.~Frau, L.~Gallot, A.~Lerda, and I.~Pesando, \emph{{Deformed N=2
  theories, generalized recursion relations and S-duality}},
  \href{http://dx.doi.org/10.1007/JHEP04(2013)039}{JHEP {\bf 1304} (2013)
  039},
\href{http://arxiv.org/abs/1302.0686}{{\tt arXiv:1302.0686 [hep-th]}}.

\bibitem{Billo:2013jba}
M.~Billo, M.~Frau, L.~Gallot, A.~Lerda, and I.~Pesando, \emph{{Modular anomaly
  equation, heat kernel and S-duality in $N=2$ theories}},
  \href{http://dx.doi.org/10.1007/JHEP11(2013)123}{JHEP {\bf 1311} (2013)
  123},
\href{http://arxiv.org/abs/1307.6648}{{\tt arXiv:1307.6648 [hep-th]}}.

\bibitem{Nemkov:2013qma}
N.~Nemkov, \emph{{S-duality as Fourier transform for arbitrary
  $\epsilon_1,\epsilon_2$}},
\href{http://arxiv.org/abs/1307.0773}{{\tt arXiv:1307.0773 [hep-th]}}.

\bibitem{Billo:2014bja}
M.~Billo, M.~Frau, F.~Fucito, A.~Lerda, J.~F. Morales, R.~Poghossian, and
  D.~Ricci~Pacifici, \emph{{Modular anomaly equations in $ \mathcal{N} =2^*$
  theories and their large-$N$ limit}},
  \href{http://dx.doi.org/10.1007/JHEP10(2014)131}{JHEP {\bf 10} (2014)  131},
\href{http://arxiv.org/abs/1406.7255}{{\tt arXiv:1406.7255 [hep-th]}}.

\bibitem{Lambert:2014fma}
N.~Lambert, D.~Orlando and S.~Reffert,
\emph{{Alpha- and Omega-Deformations from fluxes in M-Theory}},
\href{http://dx.doi.org/10.1007/JHEP11(2014)162}{JHEP {\bf 11} (2014)  162},
\href{http://arxiv.org/abs/1409.1219}{{\tt arXiv:1409.1219 [hep-th]}}.

\bibitem{Marshakov:2009kj}
A.~Marshakov, A.~Mironov, and A.~Morozov, \emph{{Zamolodchikov asymptotic
  formula and instanton expansion in N=2 SUSY N(f) = 2N(c) QCD}},
  \href{http://dx.doi.org/10.1088/1126-6708/2009/11/048}{JHEP {\bf 0911} (2009)
   048}, \href{http://arxiv.org/abs/0909.3338}{{\tt arXiv:0909.3338 [hep-th]}}.

\bibitem{KashaniPoor:2012wb}
A.-K. Kashani-Poor and J.~Troost, \emph{{The toroidal block and the genus
  expansion}}, \href{http://dx.doi.org/10.1007/JHEP03(2013)133}{JHEP {\bf 1303}
  (2013)  133},
\href{http://arxiv.org/abs/1212.0722}{{\tt arXiv:1212.0722 [hep-th]}}.

\bibitem{Kashani-Poor:2013oza}
A.-K. Kashani-Poor and J.~Troost, \emph{{Transformations of spherical blocks}},
  \href{http://dx.doi.org/10.1007/JHEP10(2013)009}{JHEP {\bf 1310} (2013)
  009},
\href{http://arxiv.org/abs/1305.7408}{{\tt arXiv:1305.7408 [hep-th]}}.

\bibitem{Kashani-Poor:2014mua}
A.-K. Kashani-Poor and J.~Troost, \emph{{Quantum geometry from the toroidal
  block}}, \href{http://dx.doi.org/10.1007/JHEP08(2014)117}{JHEP {\bf 1408}
  (2014)  117},
\href{http://arxiv.org/abs/1404.7378}{{\tt arXiv:1404.7378 [hep-th]}}.

\bibitem{Ashok:2015cba}
S.~K.~Ashok, M.~Billo, E.~Dell'Aquila, M.~Frau, A.~Lerda, and M. Raman, 
\emph{{Modular anomaly equations and S-duality in N=2 conformal SQCD}},
 \href{http://dx.doi.org/10.1007/JHEP10(2015)091}{JHEP {\bf 1510}
  (2015)  091},
\href{http://arxiv.org/abs/1507.07476}{{\tt arXiv:1507.07476 [hep-th]}}.

\bibitem{Billo}
M.~Billo, M.~Frau, F.~Fucito, A.~Lerda, and J.~F. Morales, \emph{{S-duality and
the prepotential in $\mathcal{N}=2^*$ theories (II): the non-simply laced algebras}},
\href{http://arxiv.org/abs/1507.08027}{{\tt arXiv:1507.08027 [hep-th]}} to be published in JHEP.

\bibitem{D'Hoker:1999ft}
E.~D'Hoker and D.~Phong, \emph{{Lectures on supersymmetric Yang-Mills theory
  and integrable systems}}, \href{http://arxiv.org/abs/hep-th/9912271}{{\tt
  arXiv:hep-th/9912271 [hep-th]}}.

\bibitem{Billo:2012st}
M.~Billo, M.~Frau, F.~Fucito, L.~Giacone, A.~Lerda, J.~F. Morales, and
  D.~Ricci-Pacifici, \emph{{Non-perturbative gauge/gravity correspondence in
  N=2 theories}}, \href{http://dx.doi.org/10.1007/JHEP08(2012)166}{JHEP {\bf
  1208} (2012)  166},
\href{http://arxiv.org/abs/1206.3914}{{\tt arXiv:1206.3914 [hep-th]}}.

\bibitem{Witten:1995gx}
E.~Witten, \emph{{Small instantons in string theory}},
  \href{http://dx.doi.org/10.1016/0550-3213(95)00625-7}{Nucl. Phys. {\bf B460}
  (1996)  541--559},
\href{http://arxiv.org/abs/hep-th/9511030}{{\tt arXiv:hep-th/9511030}}.

\bibitem{Douglas:1995bn}
M.~R. Douglas, \emph{{Branes within branes}},
\href{http://arxiv.org/abs/hep-th/9512077}{{\tt arXiv:hep-th/9512077}}.

\bibitem{Green:2000ke}
M.~B. Green and M.~Gutperle, \emph{{D-instanton induced interactions on a
  D3-brane}}, JHEP {\bf 02} (2000)  014,
\href{http://arxiv.org/abs/hep-th/0002011}{{\tt arXiv:hep-th/0002011}}.

\bibitem{Billo:2002hm}
M.~Billo, M.~Frau, I.~Pesando, F.~Fucito, A.~Lerda, and A.~Liccardo,
  \emph{{Classical gauge instantons from open strings}}, JHEP {\bf 02} (2003)
  045,
\href{http://arxiv.org/abs/hep-th/0211250}{{\tt arXiv:hep-th/0211250}}.

\bibitem{Billo:2006jm}
M.~Billo, M.~Frau, F.~Fucito, and A.~Lerda, \emph{{Instanton calculus in R-R
  background and the topological string}}, JHEP {\bf 11} (2006)  012,
\href{http://arxiv.org/abs/hep-th/0606013}{{\tt arXiv:hep-th/0606013}}.

\bibitem{Ennes:1999fb}
I.~P. Ennes, C.~Lozano, S.~G. Naculich, and H.~J. Schnitzer, \emph{{Elliptic
  models and M theory}},
  \href{http://dx.doi.org/10.1016/S0550-3213(00)00131-0}{Nucl. Phys. {\bf B576}
  (2000)  313--346},
\href{http://arxiv.org/abs/hep-th/9912133}{{\tt arXiv:hep-th/9912133
  [hep-th]}}.

\bibitem{Fucito:2004gi}
F.~Fucito, J.~F. Morales, and R.~Poghossian, \emph{{Instantons on quivers and
  orientifolds}}, \href{http://dx.doi.org/10.1088/1126-6708/2004/10/037}{JHEP
  {\bf 0410} (2004)  037},
\href{http://arxiv.org/abs/hep-th/0408090}{{\tt arXiv:hep-th/0408090
  [hep-th]}}.

\end{thebibliography}
\end{document}